\documentclass{emulateapj}

\def\simg{\mathrel{\rlap{\raise 0.511ex \hbox{$>$}}{\lower 0.511ex \hbox{$\sim$}}}}
\def\siml{\mathrel{\rlap{\raise 0.511ex \hbox{$<$}}{\lower 0.511ex \hbox{$\sim$}}}}

\def\eps{\varepsilon} 
\def\F0310k{{\cal F}(0.3-10\,{\rm k})}
\def\Ep{E_p}
\def\bl{\beta_L} \def\bh{\beta_H}
\def\ap{\alpha_p} \def\app{\alpha_{pp}}

\def\to{t_o} \def\tg{t_\gamma} 
\def\tp{t_p} \def\tpp{t_{pp}} 
\def\To{T_o} \def\Tg{T_\gamma} 
\def\Tp{T_p} \def\Tpp{T_{pp}} 

\def\calD{{\cal D}}
\def\n{\sl n}
\def\Gctc{\Gamma_c \theta_c}
\def\Gt{\Gamma\theta}
\def\thtc{(\theta/\theta_c)}
\def\llra{\longleftrightarrow}
\def\lra{\longrightarrow}
\def\ra{\rightarrow}

\def\h{\hspace*{-1mm}} 
\def\hh{\hspace*{-2mm}}
\def\HR{\rm HR}
\def\k{\rm k}

\def\ds{\displaystyle} 
\def\Meszaros{M\'esz\'aros~}

\begin{document}

\parskip 5pt


\title{Analysis of Two Models for the Angular Structure of the Outflows Producing the Swift/XRT "Larger-Angle Emission" of Gamma-Ray Bursts}

\author{Alin Panaitescu, Ano Neamus, Sue de Nimes, Any Author}
\affil{ISR-6, Los Alamos National Laboratory, Los Alamos, NM 87545, USA}


\begin{abstract}

 The quasi-instantaneous emission from a relativistic surface endowed with a Lorentz factor that decreases away from the outflow symmetry 
axis can naturally explain the three phases observed by Swift/XRT in GRBs and their afterglows (GRB tail, afterglow plateau and post-plateau) 
based only on the angular change of the relativistic Doppler boost across the outflow surface.
 We develop further the analytical formalism of the "Larger-Angle Emission" model for the case of "n-Exponential" outflows (where the Lorentz 
factor $\Gamma$ dependence of the angular location $\theta$ is $\Gamma \sim \exp\{-\thtc^n\}$), and compare its ability to account for the 
X-ray emission of XRT afterglows relative to that of "Power-Law" outflows ($\Gamma \sim \theta^{-g}$). 

 Power-Law outflows yield longer afterglow plateaus, followed by slower post-plateau flux decays than $\n$-Exponential outflows, features 
which may be used in identifying which type of angular structure is at work in a given afterglow.
Identifying the $\Gamma(\theta)$ angular structure that accommodates XRT light-curves is slightly complicated by that the afterglow X-ray 
light-curve is also determined by how two characteristics of the comoving-frame emission spectrum (peak-energy $E'_p$ and peak intensity $i'_p$)
change with angular location or, equivalently, with the Lorentz factor. 
 Here, we assume power-law $\Gamma$-dependences of those spectral characteristics and find that, unlike Power-Law outflows, $\n$-Exponential outflows 
cannot account for plateaus with a temporal dynamical range larger than 100 (2 dex in logarithmic space).

 To capture all the information contained in XRT afterglow measurements (0.3-10 keV unabsorbed flux and effective spectral slope), we calculate 
0.3 keV and 10 keV light-curves using a broken power-law emission spectrum of peak-energy and low- and high-energy slopes that are derived from 
the effective slope measured by XRT.
This economical peak-energy determination is found to be consistent with the results of more expensive spectral fits. 

 The angular distributions of the Lorentz factor, comoving frame peak-energy, and peak-intensity ($\Gamma (\theta), E'_p (\theta), i'_p(\theta)$) 
constrain the (yet-to-be determined) convolution of various features of the production of relativistic jets by solar-mass black-holes and of their 
propagation through the progenitor/circumburst medium, while the $E'_p (\Gamma)$ and $i'_p (\Gamma)$ dependences may constrain the GRB dissipation 
mechanism and the GRB emission process.

\end{abstract}

\vspace{2mm}
\section{\bf Introduction}


\subsection{\bf Diversity of X-ray Afterglows}

 A cursory visual inspection of the more than 1400 GRB and afterglow Swift/XRT light-curves shown at 
{\sl https://www.swift.ac.uk/xrt\_curves/allcurves.php} indicates that, for the quarter of light-curves with a sufficiently
good coverage (i.e. with a good sampling, modulo the Earth's occulation, and a long monitoring, over at least three decades 
in time) that allows the identification of all light-curve power-law segments, there are three types of afterglows:
\begin{enumerate}
\vspace*{-3mm}
\item about 90\% of light-curves display two GRB/prompt phases (one variable but roughly constant, followed by a "Tail" 
 of a sharp fall $F_\eps \sim t^{-(2-4)}$ if the time origin is close to the pulse peak) and two afterglow/delayed phases 
 (a "Plateau" of a decay slower than $F_\eps \sim t^{-3/4}$, followed by a "post-Plateau" of decay faster than $F_\eps \sim t^{-3/4}$),
\vspace*{-3mm}
\item for about 10\% of those well-monitored light-curves, the GRB with two phases is followed by an afterglow displaying
 a single decay of intermediate speed ($F_\eps \sim t^{-3/4}$) until the last measurement, although a break after the last
 measurement at 30-300 ks is still possible,
\vspace*{-3mm}
\item a minority of bursts, about few percent of the better sample, show only the first GRB and second afterglow phases, 
 i.e. the variable prompt emission is followed by a single decay steeper than $F_\eps \sim t^{-1}$ (first evidenced by Liang et al 2006).
\end{enumerate}

 The GRB Tail measured by XRT was predicted by Fenimore, Madras \& Nayakshin (1996), who studied the emission from 
the relativistic fluid moving at angles larger than $\Gamma^{-1}$ relative to the observer, with $\Gamma$ the source Lorentz 
factor. That the {\bf "Larger-Angle Emission"} ($LAE$) is dimmer than the emission from angles less than $\Gamma^{-1}$, which has 
the maximal relativistic enhancement of Doppler factor $\calD = (1-2)\,\Gamma$, thus the $LAE$ can be observed only if the
GRB source switches off sufficiently fast (faster than approximately $(t-\to)^{-3}$, with $\to$ the observer epoch when 
the GRB pulse began). The $LAE$ model was reiterated by Kumar \& Panaitescu (2000) with a correction for the Tail flux-decay, 
found to be $F_\eps \sim (t-\to)^{-(2+\beta)}$ with $\beta$ the slope of the energy spectrum at the observing energy
($F_\eps \sim \eps^{-\beta}$), and with the prediction that the $LAE$ could be used to map the angular structure of GRB outflows. 
Liang et al (2006) have used this expected decay of GRB Tails to determine the pulse beginning time $\to$, while Zhang et al (2009) 
have identified a spectral softening of the GRB Tail spectrum, which they attribute to a broken power-law spectrum whose decreasing 
peak-energy explains the observed spectral softening.

 The last phase of afterglow "normal" decay is the one observed before Swift by the Beppo-SAX mission, starting several hours after 
trigger. Prior to the modern interpretation of the X-ray emission (see below), this late X-ray emission was attributed to 
the {\bf relativistic blast-wave} (more specifically, to the {\bf external} forward-shock of that blast-wave) driven by the GRB ejecta 
into the ambient medium (\Meszaros \& Rees 1997), as was the better and more often monitored optical afterglow emission.

\vspace{2mm}
\subsection{\bf Models for X-ray Afterglow Plateaus and Chromatic X-ray Light-Curve Breaks}

	{\sl Why would one believe that GRB Tails and afterglow Plateaus have different origins} ?
 Perhaps the most interesting and consequential discovery made by XRT is the afterglow Plateau occurring at 100 s -- 30 ks
after trigger and, thus, always missed by Beppo-SAX. The sharp change in the X-ray flux evolution at the end of the GRB Tail,
from a steep decay to a slow one, accompanied almost always by a hardening of the 0.3-10 keV continuum were automatically
taken as proofs that the two phases, the GRB Tail and the afterglow Plateau, have different origins. Adding the smooth
transition from the afterglow Plateau to the normal post-Plateau phase, and the natural attribution of the afterglow post-Plateau
to the external forward-shock (above), explain the ease at which the Plateau was also attributed to that shock. 

	{\sl Why would one believe that afterglow Plateaus arise from an increase (real or apparent) in the blast-wave energy} ?
 Furthermore, that the synchrotron forward-shock emission is expected to depend most strongly on the blast-wave energy led to the 
natural assumption that the Plateau originates from a continuous {\bf injection of energy} (by arrival of delayed or slower ejecta -- 
Rees \& \Meszaros 1998 -- or by absorption of the electromagnetic radiation produced by a spinning-down magnetar -- Lyons et al 2010) 
in the blast-wave (Nousek et al 2006, Panaitescu et al 2006a, Zhang et al 2006).

 The external-shock emission being dependent on the kinetic-energy of the ejecta moving within an angle of $\Gamma^{-1}$ around the direction 
toward the observer, means that the blast-wave emission is set not only by {\bf radial} structure of the ejecta (Panaitescu \& Vestrand 2012)
(which leads to energy-injection) but also by the outflow angular structure (Granot \& Kumar 2003, Kumar \& Granot 2003, Panaitescu \& Kumar 2003).

 In a more extreme use of that factor, the blast-wave emission from an outflow seen off-axis (Eichler \& Granot 2006) could also display 
a Plateau after a sharp rise, occurring when the more energetic outflow Core becomes visible to the off-axis observer, but this model 
requires another jet pointing toward us, to account for the bright prompt emission that triggered a detector and enabled the localization that
lead to the afterglow follow-up, or observer-locations just outside the bright Core that would yield GRBs bright enough for detection
(Beniamini et al 2020).

	{\sl Why would one doubt that afterglow Plateaus are from an increasing blast-wave energy} ?
 The major issue with X-ray Plateaus arising from a variation of the ejecta kinetic-energy on the visible surface, caused either by energy-injection
or by outflow angular structure, is that all light-curve features arising from the shock's dynamics or energetics should be displayed at all 
photon energies, in contradiction with that about half of well-monitored optical and X-ray afterglows display {\bf chromatic breaks} at the end 
of the X-ray Plateau, i.e. light-curve breaks which are not seen at the same time in the optical (Panaitescu 2006b). 

       {\sl Can the external shock still account alone for afterglow chromatic light-curve breaks} ?
 The above-described limitation of the external-shock model could be circumvented if the optical and X-ray afterlow emissions were 
attributed to different shocks: {\bf reverse and forward}. The dynamics of the shocked fluids are identical because the Lorentz factor
and post-shock energy density are continuous across the contact discontinuity, therefore the decoupling of optical and X-ray light-curves 
can originate from  differences in the electron distributions with energy and in the differential dynamics of the shocks, in the case of 
synchrotron emission from fast-cooling electrons located just behind each shock.
 Then, the reverse-shock should yield a more complex electron distribution with energy because that depends on the non-monotonic histories 
of mass and Lorentz factor of the incoming ejecta, while the forward-shock's distribution is "stiffer", owing to its dependence on only the 
monotonic radial structure of the circum-burst medium (major changes in the forward-shock dynamics require excessively energetic injection
episodes in the reverse-shock).

 That complexity of the electron distribution with energy behind the reverse-shock and of this shock's dynamics could lead to diverse afterglow 
light-curves, and that is how Genet, Daigne \& Mochkovitch (2007) and Uhm \& Beloborodov (2007) succeeded in decoupling the forward-shock optical 
emission from the reverse-shock X-ray light-curve.
 If the X-ray light-curve Plateau is due mostly to the former factor (a non power-law distribution with energy), then the Plateau end 
should be accompanied by a significant spectral softening that is compatible with the post-Plateau light-curve decay steepening.
 The softening of the X-ray continuum slope by $\delta \beta = 0.2$ across the Plateau end, shown in the second reference above, is smaller
than what could be expected from the magnitude (difference between post-Plateau and Plateau flux power-law decay indices) of the accompanying 
light-curve break $\delta \alpha \simg 1$ (for comparison, a forward-shock emission spectral-break of $\delta \beta = 1/2$, crossing 
the observational band, yields a light-curve steepening of only $\delta \alpha = \delta \beta/2 = 0.25$), but still incompatible with 
the general lack of a spectral softening of the XRT X-ray continuum across the Plateau end (Figures 3-6).

     {\sl Other options: reprocessing processes}.
 Chromatic X-ray light-curve breaks could arise from a secondary process operating on a common, primary origin, and modifying the blast-wave 
emission at only one photon-energy domain, preferably in the X-rays because hundreds of articles have already attributed the optical emission 
to the forward shock. 
 In this category of reprocessing of the blast-wave emission, Panaitescu (2008) has proposed the bulk-upscattering of the optical blast-wave 
emission by an incoming, inner outflow, to the X-rays, with the more relativistic, incoming outflow also injecting energy into the external-shock. 
In this model, upscattering of the (de-beamed) optical forward-shock emission (left behind by that shock) modifies it by "showing" features 
(Lorentz factor and optical-thickness) of the incoming outflow. The model requires an almost purely leptonic incoming outflow, to ensure a
sufficiently high optical-thickness to account for the observed up-scattered X-ray flux.
 Also in this category, Shen et al (2009) have shown that scattering of the blast-wave X-ray emission by dust in the vicinity of the burst 
would yield a significant softening of the X-ray continuum during the Plateau, which is contradiction with XRT observations.

    {\sl Other options: optical and X-ray afterglows from different mechanisms}.
 The optical and X-ray afterglows could have entirely different origins and, again, it better be that the optical emission is still from 
the blast-wave.
 In this category, Ghisellini et al (2007) have identified the afterglow X-ray emission with internal shocks in a variable outflow 
produced by the "central-engine" (Rees \& \Meszaros 1994). 
 Kumar, Narayan \& Johnson (2008), and Canizzo, Troja \& Gehrels (2011) have attributed the X-ray afterglow emission to the central-engine
and have identified the afterglow Plateau with the history of the stellar envelope fall-back or to a variable accretion viscosity in 
the torus surrounding the black-hole. 

 For both these models, the central-engine must operate on a (lab-frame) timescale comparable to that of the (observer-frame) X-ray
afterglow, which can exceed 1 Ms, because switching from an the X-ray mechanism to another seems unlikely, given the generally continuous
and monotonically decreasing afterglow X-ray light-curve and the constant spectral slope of the XRT X-ray continuum after the Plateau beginning.

 The most natural model in this category seems to be that proposed by Oganesyan et al (2020), who showed that the fast-decaying 
$LAE$ from an outflow of uniform Lorentz factor turns into a slow decay if the Lorentz factor decreased on the emitting relativistic surface. 
 The reason for that much mitigated flux decay is that, when outside the $\Gamma^{-1}$ angle of maximal relativistic beaming, 
the emission-enhancement Doppler factor increases with a decreasing Lorentz factor $\Gamma (\theta)$, i.e. with location angle 
$\theta$ and, consequently, with the corresponding photon arrival-time at observer $t(\theta)$.

 The last two models are not incompatible with each other, as they focus on opposite ends of the mechanism that produces GRBs.
The fall-back central-engine model pertains to the origin of relativistic outflows whose ejection history could account for the GRB
emission and its variability while the $LAE$ model accounts for the X-ray afterglow light-curve from the ejecta that produced the last 
GRB pulse, without even specifying a dissipation and radiation mechanism, but just from the differential relativistic boost on a spherical 
surface endowed with a non-uniform Lorentz factor.

 Thus, the $LAE$ model is an {\sl add-on} to an unspecified, mysterious central-engine, and requires only a relativistic source and a mechanism 
that produces the GRB emission and operates only for the 10-100 s duration of the prompt emission, with the afterglow emission arriving 
later at observer (up to 10 days, or longer) because of the spherical curvature of the emitting surface. However, post-GRB flares 
require a central-engine operating for a longer time because their brightness seems too high for them to originate in a very bright 
"hot-spot" on an inhomogeneous spherical shell.

  {\sl LAE vs. external-shock emission}. 
 The rest of this work is dedicated to studying the $LAE$ model for two major types of outflow angular structures, after adopting it as
the mechanism that explains most naturally the diversity of XRT X-ray afterglows (Panaitescu 2020). However, it is important to note that, 
just as for the energy-injection in the external-shock interpretation for X-ray light-curve Plateaus, any mechanism that relies on dynamical
or kinematic arguments to produce light-curve features (plateaus, breaks) can yield {\sl only achromatic} features, appearing at the same
time at all photon energies. 

 Given the large brightness of the GRB prompt emission, which softens after the prompt phase and becomes X-ray emission in the $LAE$ model,
it seems more likely that the $LAE$ is dominant over the blast-wave emission at X-ray energies. Then, it seems natural to identify the X-ray 
emission with the $LAE$ and to attribute the optical emission to the external-shock for all afterglows that display chromatic X-ray 
light-curve breaks or (otherwise) decoupled X-ray and optical light-curves (with very different power-law decays). 

 The similarity of afterglows with well-coupled light-curves (same decays at any wavelength) and with achromatic breaks indicates that
both the X-ray and optical emissions arise from the same mechanism: either the $LAE$ or the external-shock with energy-injection to account 
for X-ray plateaus. However, that the $LAE$ is the softening of the bright GRB prompt emission (from hard X-ray to mid X-rays)
and that the latter is unlikely to originate in the blast-wave (one reason provided in Conclusions), strongly suggest that, for afterglows 
with well-coupled optical and X-ray light-curves, the $LAE$ should overshine the blast-wave emission not only in the X-ray but also in the optical. 

 Thus, we propose that the external shock emission is limited to only the optical emission (or, more generally, to the sub-X-ray emission)
of afterglows with decoupled light-curves and chromatic X-ray breaks.

\vspace{2mm}
\subsection{\bf The Larger-Angle Emission (LAE) Model}
\label{intro}
 
 Returning to the $LAE$ model, it should be noted that it accounts naturally for the sharp light-curve break that marks the Plateau 
beginning (as shown by Oganesyan et al 2020), the feature that tricked us to attributing the GRB and afterglow to different mechanisms. 
The other observational feature that motivated that dichotomy -- the strong spectral hardening at the Plateau beginning -- 
remains an unexplained ad-hoc feature of the $LAE$ model.\footnote{
 Measurements of the high-energy GRB spectral slope (above the GRB peak-energy $\Ep \sim 100$ keV) show a diversity
of such slopes (Preece et al 2000, Kaneko et al 2006, Poolakkil et al 2021), diversity which one could attribute to various outflow 
Lorentz factors $\Gamma$ among GRBs, via a dependence of the exponent of the shock-accelerated particle's distribution with energy
on $\Gamma$, albeit models for particle acceleration at shocks (e.g. Sironi \& Sptkovski 2011) do not indicate that such a dependence 
should exist. 
 Furthermore, the X-ray spectral slope does not display any significant evolution during the Plateau, which is the phase whose existence 
is due precisely to an evolving (decreasing) $\Gamma (\theta)$. Thus, there is no theoretical or observational reason to connect the spectral 
softening seen at the Tail-Plateau transition with the angular structure of the Lorentz factor $\Gamma$ that is essential to the $LAE$ model.}

 Despite the lack of physics (i.e. of physical processes and/or mechansims), the $LAE$ accounts naturally for 
1) the afterglow Plateaus, requiring only that the emitting surface is curved and that the Lorentz factor decreases on that surface away 
   from the symmetry axis (Oganesyan et al 2020) and
2) the above-described diversity of GRBs and afterglows (Panaitescu 2020):
\begin{enumerate}
\vspace*{-3mm}
 \item GRBs with Tails followed afterglows with Plateaus and normal decays are produced by outflows with a parameter $\Gctc \simg 1$ 
  (where $\Gamma_c$ is the axial Lorentz factor and $\theta_c$ is the angular size of the outflow Core) and an exponent of the angular 
  structure above unity ({\bf type I)}
\vspace*{-2mm}
 \item GRBs with Tails followed by afterglows with pseudo-Plateaus result for $\Gctc \simg 1$ and an angular structure exponent below unity
  {\bf type II)},
\vspace*{-2mm}
 \item GRBs without Tails followed by afterglows without Plateaus are obtained for $\Gctc \siml 1$ {\bf type III)}.
\end{enumerate}
\vspace*{-3mm}
 These phases were named according to their "visual" properties but using $LAE$ model expectations that are presented in \S\ref{PKLAE}

 To the above argument that the $LAE$ model can accommodate {\sl qualitatively} all types of X-ray afterglows in the Swift X-ray database,
one can add that any X-ray light-curve can be accommodated {\sl quantitatively} within the $LAE$ model by a large number of combinations
of the dynamical and emission angular structures in the outflow Envelope, the only limitation being in inventing such functions which, 
when "convolved" in the $LAE$ model, give the desired X-ray light-curve.
  
 Consequently, the purpose of this article is not to test the $LAE$ model (because this model does not have any significant and inherent
limitations), but rather to constrain the angular distributions of its Lorentz factor $\Gamma$ and of its comoving-frame spectral characteristics
peak-energy $E'_p$ and and peak-intensity $i'_p$. 
 For those two aims, we will develop a formalism for the $LAE$ from outflows endowed with a generalized Exponential angular structure 
(as for Power-Law outflows -- Panaitescu 2020) and we will identify differences in the afterglow emission arising from each type of structure.
 We will also calculate 0.3 keV and 10 keV specific fluxes from the 0.3-10 keV bolometric fluxes $\F0310k$ and effective 0.3-10 keV 
spectral slope, and use the two-energy light-curves to constrain the power-law angular distribution of two comoving-freame spectral 
characteristics (peak-energy $E'_p$ and peak-intensity $i'_p$).

\vspace{3mm}
\section{\bf General Features of the $LAE$ Model}

\subsection{\bf Model Assumptions}

 The following assumptions are made in our simplest version of the $LAE$ model 
\begin{enumerate}
\vspace*{-3mm}
\item the emitting surface is assumed to produce radiation for an infinitesimal (very short) time (at some radius $R$),
\vspace*{-2mm}
\item the outflow is axially-symmetric and has a quasi-uniform Core (of angular opening $\theta_c$),
\vspace*{-2mm}
\item the observer is located on the outflow symmetry axis,
\vspace*{-2mm}
\item the emitting surface angular structure is monotonic.
\end{enumerate}

\vspace*{-2mm}
 The first assumption implies that the spread in the photon-arrival time is determined solely by the spherical curvature of the emitting 
surface, and not by the thickness of the radiating fluid or the timescales on which electrons are injected/accelerated and on which they cool.
 It also and implies that the relativistic enhancement of the comoving-frame specific intensity contains the second power of the Doppler 
factor $\calD$, and not the third, because the traditional temporal contraction of a comoving-frame infinitesimal time-interval $dt'$ 
to that in the observer frame, $dt = dt'/\calD$ (not to the lab frame !), does not affect the fluence received on a (light-curve binning) 
timescale longer than $dt$, as is the case for a quasi-instantaneous emission release. 
 The price to pay for an instantaneous emission release assumption is that the radiation produced prior to the location/epoch of emission 
is "lost" and that the resulting pulse will not have a gradual rise (as a real GRB pulse) but an infinitely sharp rise. To deal with that 
limitation, GRB pulse will be fit only starting at their peaks.

 The first three assumptions above greatly simplify the calculation of light-curves because they lead to a one-to-one correspondence 
between the photon arrival-time $t$ and the angular location $\theta$ where the photon was emitted, with the Doppler enhancement 
$\calD (t)$, emission spectral characteristics $E'_p (t)$ and $i'_p (t)$, and received flux $F_\eps (t)$ being determined entirely 
by the outflow properties -- $\Gamma(\theta)$, $E'_p (\theta)$ and $i'_p (\theta)$ -- at location $\theta$.

 Off symmetry-axis observer locations will spoil that one-to-one $t(\theta)$ correspondence and will mix emissions from a range of 
locations into the flux received at a given time $t$, thus light-curve features arising from angular structure features will be spread 
in observer time $t$ and smoothed, but that is a small price to pay when considering that, for monotonic angular structures, the only
resulting sharp light-curve feature is at the GRB Tail -- afterglow Plateau transition. 

 We note that observer locations within the outflow Core should change little the resulting light-curves because the Core is assumed 
quasi-uniform. Furthermore, because the Doppler factor decreases sharply outside the Core, such observer locations lead to dimmer GRBs. 
Then, the observational bias (resulting from our propensity, predilection, and preference) for detecting and following-up the brighter 
bursts imply that selecting (for modeling) the best-monitored afterglows is equivalent to selecting outflows for which the observer 
is located within the Core. Conversely, observer locations outside the quasi-uniform Core (which produces the prompt GRB emission)
will lead to "orphan" afterglows (arising from the outflow Envelope), as shown by Ascenzi et al (2020), which may be not detected 
and localized by Swift/BAT.

\vspace{2mm}
\subsection{\bf Model Parameters}
\label{params}

 The last assumption is just a starting point in the modeling of XRT afterglows with the $LAE$ model, and enables an analytical 
treatment of the $LAE$. It means that we will employ a simple function (generalized Exponential and a power-law) for the Lorentz-factor 
$\Gamma(\theta)$ distribution and a power-law for the spectral characteristics' distributions on the emitting surface. 
 The $\Gamma(\theta)$ brings in two {\bf primary structural} model parameters: the Core parameter $\Gctc$, with $\Gamma_c$ the axial Lorentz 
factor and $\theta_c$ the Core angular opening (the $LAE$ depends on their product and not on the individual values of those parameters, 
i.e. {\sl modeling the X-ray emission with the $LAE$ cannot determine both of those two parameters}), and the power-law exponent $g$ 
(for a Power-Law) or $n$ (for a generalized Exponential).

 Furthermore, we will employ a simple power-law to quantify the distribution of the two comoving-frame spectral characteristics
peak-energy $E'_p \sim \Gamma^e$ and peak-intensity (at peak-energy) $i'_p \sim \Gamma^i$.\footnote{
 One could determine the angular variation of the spectral characteristics $E'_p$ and $i'_p$ and convert that into a $\Gamma$-dependence,
which is more relevant for the GRB mechanism, but the best-fit parameters are the same whichever parameterization is used.
 Using a model parameter instead of an independent variable for parameterizating those dependences should affect the uncertainties of 
the exponents $e$ and $i$ but, given the large $\chi^2_\nu$ of the best-fits below, such uncertainties would be meaningless underestimations.}
Thus, the simplest $LAE$ model has two {\bf secondary emission} parameters, called "secondary" because they depend on the choice of the
outflow angular structure.

 XRT observations can {\sl determine} those emission parameters only when an evolving {\sl effective} 0.3-10 keV spectral slope $\beta$ 
is measured, which indicates that the observer-frame peak-energy $\Ep$\footnote{ 
 This energy is the peak of $\eps F_\eps$ only if $\bl < 1$, which happens often during the GRB and at the Tail beginning, and if $\bh > 1$, 
which happens during the Tail and more often than not during the afterglow. In general, by "peak-energy", we mean a break-energy.}
of the XRT continuum
\begin{equation}
 F_\eps = F_p \left\{ \begin{array}{ll} (\eps/\Ep)^{-\bl} & \eps < \Ep \\ (\eps/\Ep)^{-\bh} & \Ep < \eps \end{array} \right.
\label{spectrum}
\end{equation}
is in the 0.3-10 keV band (and its evolution leads to a changing $\beta$), and only if the low-energy (below $E'_p$) and high-energy 
(above $E'_p$) spectral slopes, $\bl$ and $\bh$, do not evolve. 
Constancy of the slope $\bl$ for the Core emission (which is the GRB and its Tail) is demonstrated by the constant XRT effective slope $\beta$ 
measured by XRT around the GRB pulse peak, when $\Ep > 10$ keV. Constancy of the slope $\bh$ for the Core emission is motivated (in some cases) 
by that the continuous softening of XRT effective slope $\beta$ stops before the GRB Tail ends, which suggests that $\Ep < 0.3$ keV happens 
during the Tail. Furthermore, if the peak-energy $\Ep$ remains below the XRT band during the afterglow (produced by the Envelope),
then the (generally) unevolving XRT effective slope $\beta$ measured during the afterglow shows that the slope $\bh$ of the Envelope emission 
is also constant.

 However, XRT observations shows that, in most cases, the effective slope $\beta$ hardens at the end of the GRB Tail and start of 
the afterglow Plateau, which indicates that the {\sl high-energy slope $\bh$ is discontinuous across the Core-Envelope boundary} located 
at angle $\theta_c$. Thus, the $LAE$ model has three {\bf tertiary spectral} parameters -- $\bl$ and $\bh$ for the Core and $\bh$ 
for the Envelope.
\footnote{The fourth spectral parameter, the slope $\bl$ for the Envelope is not relevant if $\Ep < 0.3$ keV during the afterglow}

 In summary, the simplest $LAE$ model has {\sl four basic parameters that pertain to the "physics" of GRBs}: two structural parameters and 
two emission parameters, plus three spectral parameters that are relevant only to fitting the afterglow emission for determining
the four basic parameters. As shown below,
the structural parameters $\Gctc$ and indices $g/n$ are constrained primarily by the afterglow temporal milestones, while the emission 
parameters $e$ and $i$ are constrained by the differential evolution of the 0.3 keV and 10 keV specific fluxes, i.e. when the 0.3-10 keV 
effective slope $\beta$ is evolving. Unfortunately, that seems to happen mostly during the GRB Tail, thus afterglow observations do not 
determine the exponents $e$ and $i$, but set only a constraint on the combination $i + e\bh$, with the Envelope high-energy slope $\bh$ 
set by the measured effective slope $\beta$.
The three spectral parameters are constrained (in practice) by the GRB and afterglow flux decays, and (in principle) also by the
asymptotic effective spectral slope $\beta$ during the GRB and afterglow phases.

\vspace{3mm}
\section{\bf $LAE$ Model: Photon Kinematics and Light-Curves}
\label{PKLAE}

 The most important step in obtaining the $LAE$ light-curves is calculating the photon arrival-time $t(\theta)$ for a given angular
location $\theta$. That one-to-one correspondence allows then the calculation of any other quantity, for its given distribution on
the emitting surface. Considering a relativistic surface moving freely at an axially-symmetric Lorentz factor $\Gamma(\theta)$ 
until the lab-frame time $t_{lab}$ where it releases emission instantaneously, the delay in the photon arrival-time at an on-axis 
observer, relative to a photon emitted at $t_{lab} = 0$ from the outflow origin, and caused by the spherical curvature of the emitting 
surface, is $t(\theta) = t_{lab}[1 - v(\theta) \cos \theta]$, which can be cast as
\begin{equation}
  ({\rm Photon-Kinematics}) \quad \left(\frac{\Gamma_c}{\Gamma(\theta)}\right)^2 + (\Gamma_c \theta)^2 = \frac{t}{\to} 
\label{pke}
\end{equation}
where 
\begin{equation}
 \to = t_{lab} [1-v(\theta=0)] \simeq \frac{t_{lab}}{2\,\Gamma^2_c}
\label{to}
\end{equation}
is the arrival-time of the photons emitted by the fluid on the symmetry axis (moving precisley toward the observer at Lorentz factor $\Gamma_c$), 
at lab-frame time $t_{lab} = R/v(\theta=0)$ (for a source radius $R$), both measured since the outflow ejection.\footnote{
 In the $LAE$ model, the afterglow emission is produced by the last GRB pulse, thus $\to$ could be well after the trigger epoch, 
especially when the GRB localizing instrument triggered on a precursor. Furthermore, in real GRBs, the pulse emission may be produced 
before the epoch $t_{lab}$ and the pulse may start before $\to$, i.e. $\to$ should be after the arrival of the first photons from 
the last GRB pulse, with the pulse-peak epoch being, most likely, the best choice.}
In Equation (\ref{pke}), $\theta \ll 1$ and $\Gamma \gg 1$ were assumed.

 The first term in the left-hand side of the "photon kinematics" (PK) Equation (\ref{pke}) arises from the angular structure 
of the emitting surface and {\sl is dominant when $\Gt < 1$}, in which case 
\begin{equation}
 ({\rm non-dynamical}) \quad \Gamma (t) = \Gamma_c \left( \frac{\ds t}{\ds \to}  \right)^{-1/2}
\label{Gamma}
\end{equation}
{\sl independent of the outflow structure}, although this conclusion is valid only if $\Gamma \ll \Gamma_c$, i.e. sufficiently far from
the outflow symmetry axis which, for the structures considered here, means outside the quasi-uniform outflow Core ($\theta > \theta_c$).

 Furthermore, the second term in the $lhs$ is from the spherical curvature of the emitting surface and {\sl is dominant when $\Gt > 1$}, 
in which case
\begin{equation}
  \theta (t) = \frac{\ds 1}{\ds \Gamma_c} \left( \frac{\ds t}{\ds \to}  \right)^{1/2}
\label{theta}
\end{equation}
{\sl independent of the outflow structure}.

 To proceed, one must consider a specific angular structure $\Gamma(\theta)$ and solve the PK Equation (\ref{pke}) for $\theta(t)$, 
to arrive at $\Gamma(t)$ and to the time-dependence of all other relevant quantities. That is only for analytical calculations because 
the PK equation provides $t(\theta)$, which suffices for numerical calculations, where a specified angular structure $\Gamma(\theta)$ yields 
immediately the temporal evolution $\Gamma [t(\theta)]$.

 The first structure considered below is a generalized exponential, the second is a power-law. Oganesyan et al (2020) and Ascenzi et al 
(2020) have calculated 
numerically the light-curves from Gaussian outflows, Oganesyan et al (2020) have focused their numerical calculations mostly to Power-Law 
structures, while Panaitescu (2020) has provided both an analytical and numerical treatment of Power-Law outflows. To compare both
types of outflows, we develop here an analytical framework for generalized-Exponential outflows and complete the existing one for Power-Law
structures. Then, numerical light-curves obtained for both types of outflows will be fit to XRT aferglows of different characteristics.

\vspace{2mm}
\subsection{\bf Generalized-Exponential Outflows}
\label{Exp}

 After noting that the terminology "power-law exponential" can be ambiguous, we define an $\n$-Exponential outflow by
\begin{equation}
 \Gamma (\theta) = \Gamma_c \; e^{-\frac{\ds 1}{\ds n} \left(\frac{\ds \theta}{\ds \theta_c}\right)^n}
\label{nExpo}
\end{equation}
A simple exponential corresponds to $n=1$, a Gaussian (Nature's favorite distribution) corresponds to $n=2$.
The $n$ appears in the denominator of the exponential for a simpler result below and to allow the identification of $\theta_c$ 
with the $1\sigma$ dispersion of a Gaussian. It is also has the effect of taming that fast decrease of $\Gamma$ in the outflow 
Envelope ($\theta > \theta_c$), with the side-effect that it makes for an anodine decrease of $\Gamma$ in the Core ($\theta < \theta_c$). 
Put together, the presence (or absence) of $n$ in the exponent's denominator does not significantly the ability of an $\n$-Exponential
outflow to accommodate XRT afterglows, and the best-fits shown below did not improve when the index $n$ was removed from the denominator.

 For this $\Gamma(\theta)$ angular exponential, the (PK) Equation (\ref{pke}) becomes
\begin{equation}
 e^{\frac{2}{n}x^n} + (\Gctc)^2 x^2  = \frac{t}{\to} \;, \quad x \equiv \frac{\theta}{\theta_c}
\label{pkeExp}
\end{equation}
One can derive approximate solutions $x(t)$ by studying where each term (from the angular structure or from the spherical curvature) 
in the left-hand side above is dominant.
 To that end, we note that the curves $e^{2x^n/n}$ and $(\Gctc)^2 x^2$ are tangent for $\Gctc = e^{1/n}$ at $x_o = 1$ 
(i.e. at the Core-Envelope boundary at $\theta=\theta_c$), which leads to that these two curves intersect only if $\Gctc > e^{1/n}$,
and to that $e^{x^n/n} > (\Gctc) x$ for any $x$ if $\Gctc < e^{1/n}$, and for $x < x_\gamma$ \& $x > x_{pp}$ if $\Gctc > e^{1/n}$, 
with $x_\gamma$ and $x_{pp}$ solutions to equation $e^{x^n/n} = (\Gctc) x$ that will be determined (approximately).

\vspace*{2mm}
\subsubsection{Temporal Milestones (Light-Curve Breaks)}

 First, we define $y = (\Gctc) x$ and note that 
\begin{equation}
 \quad e^{\frac{1}{n}x^n} = (\Gctc) x \quad \llra \quad  y^n = n (\Gctc)^n \ln y
\label{yton}
\end{equation}
for $\Gctc > e^{1/n}$.
Owing to the fast increase of $y^n$, one expects the solution to the last equation to be slightly above unity and defining 
$y_{min}=1+\eps$ with $\eps \ll 1$ leads to $\eps = 1/[n(\Gctc)^n]$, thus $\eps \ll 1$ only for $n (\Gctc)^n > n e \gg 1$, 
which is approximately satisfied for $n > 1$.\footnote{
 Because $e$ is not the electron charge but Euler's number $e \simeq 3$, then $e \gg 1$ does not require a big stretch of imagination}
 Thus, $x_\gamma \simg (\Gctc)^{-1}$ should be a decent approximation.

 The meaning of $x_\gamma$ is shown by that $\theta(x_\gamma) \simg \Gamma_c^{-1}$. Taking into account that $\Gamma (\theta=1/\Gamma_c) 
\siml \Gamma_c$, it follows that $(\Gt) (x_\gamma) \simeq 1$, thus the near-axis region of maximal Doppler boost (where $\Gt < 1$), 
corresponding to the prompt GRB emission, is defined by $x < x_\gamma$. Further, by replacing $x_\gamma$ in Equation (\ref{pkeExp}), 
one obtains that the GRB phase should last for a duration
\begin{equation}
 \tg \equiv t(x_\gamma) \simg 2\, \to
\label{tgrb}
\end{equation}
for $\Gctc > e^{1/n}$,
i.e. {\sl the GRB pulse duration $\tg$}, defined as the phase of maximal relativistic enhancement ($\cal D \simg \Gamma_c$), 
{\sl is comparable to the pulse-peak time $\to$} 
measured since the relativistic surface was ejected. However, that ejection epoch is not easily measurable from the pulse light-curve and, 
as described above, the pulse rise is not captured in our $LAE$ model because the emission is assumed to be instantaneous at $\to$). 
Instead, as shown below, the GRB pulse duration $\tg$ can be estimated from the $LAE$-model expected flux-decrease by about 
$2^{\beta + 2} \simeq 10$ from the pulse peak-flux, where $\beta$ is the spectral slope at the observing energy.

 Equation (\ref{yton}) can also be solved for the higher solution $x_{pp}$ after noting that its second form has the interesting 
feature that it converges iteratively quite fast to the actual solution no matter how far one starts with an initial $y_o$ guess in 
the rihgt-hand side, for any reasonable index $n \in [1,4]$, and for any reasonable Core parameter $\Gctc \in [1,10]$:
several iterations are sufficient to arrive within 1\% of the solution. Thus, solving Equation (\ref{yton}) numerically is easy,
but solving it analytically requires a slightly tricky identification of an average coefficient $k$ for the solution 
$y_{max} = k (\Gctc) (\ln \Gctc)^{1/n}$ that brings the analytical solution close to the numerical one for the above ranges
of $n$ and $\Gctc$. We find that that coefficient is 1.6, leading to $x_{pp} \simeq 1.6\,(\ln \Gctc)^{1/n}$.

 The meaning of $x_{pp}$ (as well as that of $x_\gamma$) can be understood after noting that 
\begin{equation}
 \Gt = (\Gctc)\; x\; e^{-\frac{1}{n}x^n}    \left\{ \begin{array}{ll} 
    < 1 & x < x_\gamma \\ = 1 & x = x_\gamma \\ > 1 & x_\gamma < x < x_{pp} \\ = 1 & x = x_{pp} \\ < 1 & x_{pp} < x
    \end{array} \right.
\label{Gt}
\end{equation}
 The value of $\Gt$ below (above) unity defines when the emission is beamed toward (away from) the observer, or enhanced (diminished)
by the relativistic motion of the source, and that is quantified by the Doppler factor
\begin{equation}
 \calD \simeq \frac{2 \Gamma}{1 + (\Gt)^2} \simeq \left\{ \begin{array}{ll} 
         2\Gamma  & \theta \ll \Gamma^{-1} \\ 2/(\Gamma \theta^2) & \theta \gg \Gamma^{-1} \end{array} \right.
\label{doppler}
\end{equation}

 Consequently, $x < x_\gamma \simeq (\Gctc)^{-1}$ corresponds to $\theta \siml \Gamma_c^{-1}$, $\Gamma \siml \Gamma_c$, $\Gt < 1$
(radiation enhanced/beamed toward the observer), and to a maximal Doppler factor $\calD \simeq 2\Gamma \siml 2\Gamma_c$, having a weak 
time-dependence. Thus, $x < x_\gamma$ corresponds to the bright and steady GRB emission at $\to < t < \tg \simeq 2\, \to$ (as discussed above).
 Furthermore, $x > x_{pp}$ corresponds to $\Gt < 1$ (radiation enhanced/beamed toward the observer), and to a maximized Doppler 
factor $\calD \simeq 2\Gamma$, having a strong time-dependence. Thus, $x_{pp} < x$ corresponds to the last phase of the afterglow, 
the return to the "normal" flux-decay (i.e. to the "post-Plateau" phase) occurring after the end of the Plateau at
\begin{equation}
 \tpp \equiv t(x_{pp}) \simeq 5.4\, (\ln \Gctc)^{2/n} (\Gctc)^2 \to
\label{tppExpo}
\end{equation}
for $\Gctc > e^{1/n}$.

 The remaining regime, defined by $x_\gamma < x < x_{pp}$, corresponds to $\Gt > 1$ (radiation diminished/beamed away from the observer),
during which the Doppler factor $\calD \simeq 2/(\Gamma \theta^2)$ evolves fast. This regime corresponds to the "GRB-Tail" (when $\Gt$
increases in time) and the afterglow Plateau (when $\Gt$ decreases in time), as will be shown below, separated by the epoch when
$\Gt$ reaches a maximum. Derivating Equation (\ref{Gt}) shows that that maximum is reached when $x=1$, i.e. at $\theta=\theta_c$,
which is the Core-Envelope boundary (the index $n$ was put in the $n$-Exponential outflow definition in Eq (\ref{nExpo}) only to reach 
this simple result), and at a time that Equation (\ref{pkeExp}) shows to be 
\begin{equation}
 \tp \equiv t(x=1) = [(\Gctc)^2 + e^{2/n}] \to
\label{tpExpo}
\end{equation}
for any $\Gctc$.

 Equations (\ref{tgrb}) (\ref{tppExpo}) and (\ref{tpExpo}) show that the temporal milestones of the $LAE$ emission -- the GRB/GRB-Tail, 
the Tail/Plateau,
and the Plateau/post-Plateau transition epochs -- depend on the structural parameters $n$ and $\Gctc$. Because the flux during each phase 
is independent of the parameters axial Lorentz factor $\Gamma_c$ and Core angular opening $\theta_c$, it follows that the $LAE$ light-curves 
depend on the product $\Gctc$ and not on the individual values of $\Gamma_c$ and $\theta_c$, a conclusion that will also be true for 
Power-Law outflows (below). Thus, {\sl the structural parameters $\Gamma_c$ and $\theta_c$ are degenerate}, in the sense that Swift 
measurements of the $LAE$ cannot determine them individually.

 Returning to solving Equation (\ref{pkeExp}) in asymptotic regimes, we reiterate the previous conclusion that the angular structure 
term $e^{x^n/n}$ is dominant for any $x$ {\bf if $\Gctc < e^{1/n}$}, in which case $\Gt < 1$ (also from Equation \ref{Gt}) for any $x$, 
i.e. solutions $x_\gamma$ and $x_{pp}$ do not exist and the maximal value of $\Gt$ (reached at $x=1$) is $\max(\Gt) = \Gctc/e^{1/n} < 1$. 
In this case, $\tg = t(x_\gamma)$ and $\tpp = t(x_{pp})$ do not exist either, and the GRB/afterglow has only two phases, separated by 
$\tp = t(x=1)$: the "GRB phase", characterized by an increasing $\Gt < 1$ and the afterglow "normal-decay phase" (or the post-Plateau 
of an otherwise non-existing Plateau !), characterized by a decreasing $\Gt < 1$.

 {\bf For $\Gctc > e^{1/n}$}, $x_\gamma/x_{pp}$ and $\tg/\tpp$ exist, and the angular structure term $e^{x^n/n}$ is dominant and 
$\Gt < 1$ both before $\tg$ and after $\tpp$. Consequently, the GRB phase ends at $\tg$, being followed by a GRB-Tail (defined by 
an increasing $\Gt > 1$), which ends at $\tp$ (when $\max(\Gt) = \Gctc/e^{1/n} > 1$), being followed by the afterglow Plateau 
(characterized by a decreasing $\Gt > 1$), which ends at $\tpp$, followed by the afterglow post-Plateau. 

 Below, we show the solutions to the (PK) Equation (\ref{pkeExp}) for these four regimes, and for $\Gctc > e^{1/n}$, with the understanding
that, for $\Gctc < e^{1/n}$, only the first and last phases exist.

\vspace{2mm}
\subsubsection{GRB Phase $t < \tg$ (increasing $\Gt < 1$)}

 Although the angular structure term $e^{x^n/n}$ is dominant over the spherical structure term $(\Gctc)x$, it would be foolish to ignore 
the latter. Instead, the first order approximation $e^{x^n/n} \simeq 1 + x^n/n$ (which is good for $x \ll 1$, but is applied here for 
$x < x_\gamma = (\Gctc)^{-1} \siml 1$) leads to
\begin{equation}
 \frac{2}{n} x^n + (\Gctc)^2 x^2  = \frac{t}{\to} - 1
\end{equation}
thus the power-law term can be ignored only if $n < 2$. In the limit $x \ll 1$, the approximate solution is
\begin{equation}
 \theta (t) = x \theta_c \simeq \left\{ \begin{array}{ll} 
          \theta_c \left( \frac{\ds n}{\ds 2} \frac{\ds t-\to}{\ds \to} \right)^{1/n} & n < 2 \\
          \theta_c \frac{\ds (t/\to - 1)^{1/2}} {\ds \left[ (\Gctc)^2+1 \right]^{1/2} } & n=2 \\
          \Gamma_c^{-1} \left( \frac{\ds t}{\ds \to} -1 \right)^{1/2} & 2 < n 
     \end{array} \right.
\end{equation}
 Note that this solution depends on the angular structure index $n$ only if $n < 2$.

 One can also calculate the corresponding Lorentz factor "evolution"
\begin{equation}
 \Gamma (t) \simeq \Gamma_c \left\{ \begin{array}{ll} e^{- \frac{\ds 1}{\ds 2}\left(  \frac{\ds t}{\ds \to} -1 \right)} & n < 2 \\ 
           e^{- \frac{\ds 1}{\ds n} \left( \frac{\ds t - \to}{\ds \tp} \right)^{n/2}}  & 2 \leq n
     \end{array} \right. 
\end{equation}
where $\tp \simeq (\Gctc)^2 \to$ was used for the second line, which is appropriate for $\Gctc \gg e^{1/n}$ (Equation \ref{tpExpo}).

\vspace{2mm}
\subsubsection{GRB-Tail $\tg < t < \tp$ (increasing $\Gt > 1$) and Afterglow Plateau $\tp < t < \tpp$ (decreasing $\Gt > 1$)}

 When the spherical curvature term $(\Gctc) x$ is dominant over the angular structure term $e^{x^n/n}$, the solution to Equation 
(\ref{pkeExp}) is independent of the angular structure ans is given in Equation (\ref{theta}), which leads to
\begin{equation}
 \Gamma (t) = \Gamma_c\, e^{-\frac{\ds (t/\to)^{n/2}}{\ds n(\Gctc)^n}} \simeq 
  \Gamma_c\, e^{- \frac{\ds 1}{\ds n} \left( \frac{\ds t}{\ds \tp} \right)^{n/2}}
\end{equation}
where $\Gctc \gg e^{1/n}$ was assumed for the latter result.

\vspace{2mm}
\subsubsection{Afterglow Post-Plateau $\tpp < t$ (decreasing $\Gt < 1$)}

 Here, we return to the angular structure term $e^{x^n/n}$ being dominant over the spherical curvature term $(\Gctc)x$, which leads to
\begin{equation}
 \theta (t) = \theta_c \left( \frac{\ds n}{\ds 2} \ln \frac{\ds t}{\ds \to} \right)^{1/n} 
\end{equation}
 and the Lorentz factor, given in Equation (\ref{Gamma}), is independent on the angular structure.

\vspace{2mm}
\subsubsection{Light-Curves and Power-Law Flux-Decay Indices}

 Once $\Gamma(t)$ on the emitting surface is known, the GRB and afterglow light-curves can be calculated (equation 19 of Panaitescu 2020)
\begin{equation}
 F_\eps \sim D^2 i'\left( \frac{\eps}{D} \right) \frac{d\theta^2}{dt} \;, \quad 
 i'(\eps') = i'_p (\theta) \left( \frac{\eps'}{E'_p(\theta)} \right)^{-\beta}
\label{lc}
\end{equation}
provided that the characteristics of the comoving-frame spectrum are parameterized :
\begin{equation}
 i'_p (\theta) \sim \Gamma^i \;, \; E'_p (\theta) \sim \Gamma^e \lra \i'(\eps') \sim \Gamma^{i + e\beta} (\eps')^{-\beta}
\label{ipEp}
\end{equation}
with $\beta$ the spectral slope at observing photon energy $\eps = \calD \eps'$.

 These lead to $LAE$ light-curve
\begin{equation}
 F_\eps (t) \sim \left[ D^{2+\beta} \Gamma^{i + e\beta} \frac{d\theta^2}{dt} \right]_t \; \eps^{-\beta}
\label{LAEflux}
\end{equation}
from where the power-law flux-decay index
\begin{equation}
 \alpha = - \frac{\ds d\ln F_\eps}{\ds d\ln t} \quad \left( F_\eps \sim t^{-\alpha} \right)
\label{alpha}
\end{equation}
can be calculated.

 Results for all four phases above are listed in Table 1.

\vspace{2mm}
\subsection{\bf Power-Law Outflows}
\label{PL}

 For an outflow having a {\sl uniform Core} of Lorentz factor $\Gamma_c$ and a {\sl Power-Law Envelope} of index $g$
\begin{equation}
  \Gamma (\theta) = \Gamma_c  \left( \frac{\theta}{\theta_c} \right)^{-g} \;, \quad 
    g \left\{ \begin{array}{ll} = 0 & \theta < \theta_c \\  > 0 & \theta_c < \theta 
                        \end{array} \right.
\label{plaw}
\end{equation}
the PK Equation (\ref{pke}) becomes
\begin{equation}
 x^{2g} + (\Gctc)^2 x^2 = \frac{t}{\to} 
\label{pkePLaw}
\end{equation}

\vspace{2mm}
\subsubsection{GRB and Tail (increasing $\Gt$) from the Core ($\theta < \theta_c$)}
\label{grbtail}

 For the uniform outflow Core ($x < 1$), we have $g=0$, $\Gamma = \Gamma_c$, and Equation (\ref{pkePLaw}) becomes
\begin{equation}
 (\Gctc)^2 x^2 = \frac{t}{\to} - 1
\label{pkeCore}
\end{equation}
containing only the spherical curvature term, with the simple solution 
\begin{equation}
  \theta (t) = \frac{\ds 1}{\ds \Gamma_c} \left( \frac{\ds t}{\ds \to} - 1 \right)^{1/2}
\end{equation}

 The increasing $\Gamma\theta = (\Gctc) x$ reaches unity at $\theta_\gamma = \Gamma_c^{-1}$, which defines the end of the GRB phase 
at the epoch $\tg$ given in Equation (\ref{tgrb}), but that happens within the Core only for $\theta_\gamma < \theta_c$, i.e. for 
$\Gctc > 1$, which is similar to the condition $\Gctc > e^{1/n}$ encountered for an $\n$-Exponential structure.

 Thus, only for $\Gctc > 1$ is the Core GRB emission (defined by an increasing $\Gt < 1$) followed by a GRB-Tail (defined by an 
increasing $\Gt > 1$), with the latter lasting until $\Gt$ reaches a maximum, at the Core-Envelope boundary ($x=1$, $\theta=\theta_c$), 
where $\max (\Gt)= \Gctc > 1$. The end of the GRB-Tail (and beginning of the afterglow Plateau) can be calculated 
from Equation (\ref{pkeCore}):
\begin{equation}
 \tp \equiv t(x=1) = [(\Gctc)^2 + 1] \to
\label{tpPLaw}
\end{equation}
which is similar to the result for an $n$-Eponential structure (Equation \ref{tpExpo}).

\vspace{2mm}
\subsubsection{Envelope ($\theta_c < \theta$) Emission}

 The two afterglow phases are separated by $\Gt = (\Gctc) x^{1-g}$ (the ratio of the spherical curvature to the angular structure term
in Equation \ref{pkePLaw}) reaching unity at
\begin{equation}
 x_{pp} = (\Gctc)^{\frac{1}{g-1}}  > 1 \; {\rm for} \; \left\{ \begin{array}{ll} 
        g < 1 \;,\; \Gctc < 1 \\ g > 1 \;,\; \Gctc > 1 \end{array} \right.
\end{equation}
For remaining case, $x_{pp} < 1$, thus $\Gt$ does not reach unity in the Envelope and the afterglow will have only one phase.

 From Equation (\ref{pkePLaw}), $\Gt (x_{pp}) = 1$ occurs at 
\begin{equation}
 \tpp \equiv t(x_{pp}) \simeq  2\,(\Gctc)^{\frac{2g}{g-1}} \to
\label{tppPLaw}
\end{equation}
 which represents the Plateau end and the post-Plateau start, according to the adopted definitions for those phases.

\vspace{2mm}
\subsubsection{Post-GRB Phases}

\vspace{2mm}
\centerline{\sl i)\, $\Gctc < 1$, $g < 1$ $\lra$ two post-GRB phases}

 For $\Gctc < 1$, the Core emission satisfies the condition of "increasing $\Gt < 1$", which defines the GRB phase, and that lasts until 
$\tp$, when the Core edge is reached. After the Envelope becomes visible at $\tp$, the condition of "increasing $\Gt < 1$" remains 
satisfied until $\tpp$, thus the Envelope emission is relativistically enhanced toward the observer (as for the GRB and post-Plateau)
and displays a slow decay (as for the Plateau) (see below). After $\tpp$, the Envelope emission satisfies the condition "increasing 
$\Gt > 1$" (which defines the Tail) and displays a very fast decay (as for the Tail) (see below). Thus, the Envelope emission displays 
visual features and satisfies conditions that are characteristic of all possible four phases !   

 The two phases for the Envelope emission are (with the first being very ambiguous):
\begin{equation}
 \theta = \left\{ \begin{array}{llll}  
  \theta_c \left( \frac{\ds t}{\ds \to} \right)^{1/2g}     & \tp < t < \tpp & ({\rm ??}) \\
  \Gamma_c^{-1} \left( \frac{\ds t}{\ds \to} \right)^{1/2} & \tpp < t       & ({\rm Tail })\\
 \end{array} \right.
\label{theta1}
\end{equation}
\begin{equation}
 \Gamma = \Gamma_c \left\{ \begin{array}{llll}  
  \left( \frac{\ds t}{\ds \to} \right)^{-1/2}           & \tp < t < \tpp  \\
  (\Gctc)^g \left( \frac{\ds t}{\ds \to} \right)^{-g/2} & \tpp < t        \\
 \end{array} \right.
\end{equation}
\begin{equation}
 \alpha = \left\{ \begin{array}{llll}  
   2 + \frac{\ds \beta}{\ds 2} - \frac{\ds 1}{\ds g}   & \tp < t < \tpp  \\
  (2 + \beta) \left( 1 - \frac{\ds g}{\ds 2} \right)   & \tpp < t        \\
 \end{array} \right.
\label{alpha1}
\end{equation}

 The power-law flux-decay indices above are for non-evolving spectral characteristics ($i=e=0$ in Equation \ref{ipEp}).

\vspace{4mm}
\centerline{\sl ii)\, $\Gctc < 1$, $g \geq 1$ $\lra$ one post-GRB phase}

 After the Envelope becomes visible at $\tp$, the condition of a "decreasing $\Gt < 1$" (as for a post-Plateau) is satisfied indefinitely,
but the Envelope emission displays a decay faster than usually observed for post-Plateaus. 
The relevant results are those on the first line of Equations (\ref{theta1})-(\ref{alpha1}).

\vspace{4mm}
\centerline{\sl iii)\, $\Gctc > 1$, $g \leq 1$ $\lra$ two post-GRB phases}

 As discussed in \S\ref{grbtail}, for $\Gctc > 1$, the Core produces a Tail (defined by an "increasing $\Gt > 1$") after $\tg$ and until $\tp$.
The relevant results are listed in Table 1 for the Tail.

 For $g \leq 1$, that condition remains satisfied by the Envelope emission after $\tp$, but the flux decay is slower.
The relevant results for this ambiguous Envelope phase are as for the second line of Equations (\ref{theta1})-(\ref{alpha1}).
Note that, for $\Gctc \gg 1$ we have $\tp \simeq (\Gctc)^2 \to$, and those results can be re-written as
\begin{equation}
 \theta = \theta_c \left( \frac{\ds t}{\ds \tp} \right)^{1/2} ,\; \Gamma = \Gamma_c \left( \frac{\ds t}{\ds \tp} \right)^{-g/2} (\tp < t) 
\end{equation}

\vspace{2mm}
\centerline{\sl iv)\, $\Gctc > 1$, $g >1 $ $\lra$ three post-GRB phases}

 In this case, all four phases identified for an $\n$-Exponential outflow are encountered, and they satisfy their defining conditions:
a Tail with an "increasing $\Gt > 1$", an afterglow Plateau with a "decreasing $\Gt > 1$", and a post-Plateau with a "decreasing $\Gt < 1$".
The relevant results for the Envelope are those given in Equations (\ref{theta1})-(\ref{alpha1}), but with their order reversed.

\vspace{2mm}
\subsection{\bf Power-Law vs. $\n$-Exponential Outflows }

 The important {\sl analytical} results for generalized-Exponential and Power-Law outflows are contrasted in Table 1 and in Figures 1 and 2.

 As discussed in \S\ref{PKLAE}, the solution $\theta(t)$ to the PK Equation (\ref{pke}) is independent of the outflow structure 
if the spherical curvature term is dominant (Equation \ref{theta}), i.e. if $\Gt > 1$, which happens during the Tail and Plateau.
Furthermore, the resulting Lorentz factor evolution (in a non-dynamical sense) $\Gamma(t)$ (Equation \ref{Gamma}) is independent 
of the outflow structure when the angular structure term is dominant, which happens during the GRB and the post-Plateau. 
 These general features are indicated in Table 1. 

 Table 1 also compares the power-law flux-decay indices $F_\eps \sim t^{-\alpha} \eps^{-\beta}$, which depends on the spectral slope $\beta$
at photon energy $\eps$, on the structural indices $n$ or $g$, and on the evolution of the comoving-frame spectral characteristics 
(Equation \ref{ipEp}).
 For an $\n$-Exponential outflow, the power-law decay indices $\alpha$ are "local" because they depend on time $t$ (i.e. the flux-decay
is not a power-law in time). 
 For a Power-Law outflow, the index $\alpha$ is asymptotic, i.e. the light-curve is a power-law in time far enough from the breaks at 
GRB end ($\tg$), Plateau beginning ($\tp$), and Plateau end ($\tpp$).
 Only the GRB and the post-Plateau phases display similar flux-decay indices, with a larger difference for the latter, where
$\app^{(Exp)} - \app^{(PL)} (t \gg \to) = 1/g$ indicates that the post-Plateau flux-decay of any generlized-Exponential outflow
is faster than that of a Power-Law outflow.  

 From \S\ref{Exp} and \S\ref{PL} results a potentially important discriminator: $\n$-Exponential outflows yield two or four of the phases 
listed in Table 1 and account only in part for the diversity displayed by XRT afterglows (\S\ref{intro}), while Power-Law outflows 
yield two, three, or four of those phases and, thus, explain the entire afterglow diversity.

 A comparison of the {\sl numerical} results for each type of angular structure is presented in \S\ref{NumFits}.

\vspace{2mm}
\subsection{\bf Determination of $LAE$ Model Structural Parameters from the Timing of Afterglow Light-Curve Breaks} 

 The Plateau beginning epoch $\tp$ given in Equations (\ref{tpExpo}) and (\ref{tpPLaw}) is about the same for both types of angular structures,
and it provides a way of estimating the $\Gctc$ structural parameter:
\begin{equation}
 \Gctc \simeq \left( \frac{\ds \tp}{\ds \to} - 1 \right)^{1/2} 
\label{Gctc}
\end{equation}
Combining with Equations (\ref{tppExpo}) and (\ref{tppPLaw}), allows a dtermination of the structural parameter $n$ or $g$:
\begin{equation}
 n \simeq 2 \frac{\ds \ln \left[ \frac{1}{2} \ln \left(\frac{\tp}{\to}-1\right) \right] }{ \ln \left(\frac{\ds \tpp}{\ds 5.4 \tp} \right) } \;,
 \quad
 g = 1 + \frac{ \ds \ln \left( \frac{\tp}{\to}-1 \right) }{ \frac{\ds \tpp}{\ds 2\tp}  }
\label{ng}
\end{equation}
with the first equality only for $\Gctc > e^{1/n}$ and the last only for $\Gctc > 1$.

 However, use of Equation (\ref{ng}) to determine the structural parameters is a bit complicated by that the ejection-to-emission duration $\to$ 
and the Plateau end epoch $\tpp$ are not easy to read from an X-ray light-curve. Additionally, Equation (\ref{nExpo}) is an approximation and so 
is the above calculation of the index $n$.

 As discussed in \S\ref{PKLAE}, the observer-frame duration $\to$ corresponding to the lab-frame duration $t_{lab}$ (from outflow ejection
and its instantaneous production of the GRB emission) can be much shorter than the time since GRB trigger, and can be longer than rise time
of the GRB pulse.
The GRB duration $\tg \simg 2\to$, (Equation \ref{tgrb}) may help in estimating $\to$ if the Tail decay of index $\alpha = 2 + \beta$, 
expected after $\tg$, is blithely extrapolated to the GRB phase before $\tg$, so that, from $\to$ to $\tg$, the flux would decay by a 
factor $(\tg/\to)^{2+\beta} = 2^{2+\beta} \simeq 10$. This appears to be a good approximation, leading to the rule-of-thumb that $\to$ 
is the duration over which the GRB-pulse peak-flux decreases by a factor 10.

 The Plateau beginning at $\tp$ is easier to measure directly from X-ray light-curves because it appears as a sharper break but $\tpp$ 
is more difficult because the Plateau to post-Plateau transition is much slower and the break less sharp. The advantage is that both
$\Gctc$ and $\tpp$ appear in a logarithm, thus the error in estimating the structural index $n$ or $g$ may be small.

 The structural exponents $n$ and $g$ also determine the Plateau and post-Plateau flux-decay indices $\ap$ and $\app$ (Table 1), 
which also depend on the exponents $e$ and $i$ for the evolution of the comoving-frame spectral characteristics, thus the measured flux-decay 
indices {\sl constrain} the structural exponents {\sl only when} the decay indices are determined by the combination $i + e\beta$ 
(i.e. only when the XRT continuum is a power-law) because, in the opposite case (when the peak-energy $\Ep$ is in the XRT band
and the X-ray continuum is a broken power-law), the decay indices $\ap$ and $\app$ determine the spectral indices $i$ and $\beta$.

 Put together, one could use Equations (\ref{Gctc}) and (\ref{ng}) to estimate the structural parameters from the timing (and decay) of X-ray 
light-curves, but that may not be a good-enough substitute for numerical fits.

\vspace{3mm}
\section{\bf Application to Swift X-ray Afterglows}

\vspace{2mm}
\subsection{\bf Calculation of XRT X-ray Flux Densities}
\label{XRT}

 To capture all the information in XRT measurements, one could fit simultaneously the 0.3-10 keV bolometric flux $\F0310k$ and 
the reported 0.3-10 keV effective slope, but that would lead to a weird hybrid "metric" that combines quantities that have a different meaning.
Instead, one should convert the available $\F0310k$ and $\beta$ to fluxes at two photon energies, as far from each other as possible,
i.e. at 0.3 keV and 10 keV, with the former requiring the use of the {\sl unabsorbed flux} $\F0310k$ available at the {\sl Swift Burst 
Analyser repository} (Evans et al 2010).

 {\sl Appendix A} shows how we calculate afterglow flux densities at two photon energies from the available XRT data so that the resulting
light-curves contain all the information existing in that data, including any spectral evolution quantified by the effective spectral slope.

\vspace{2mm}
\subsection{\bf Numerical Calculation of $LAE$ Light-Curves}
\label{numerics}

 For numerical calculations of the $LAE$ 
\begin{enumerate}
\vspace*{-3mm}
\item Equation (\ref{pke}) is used to determine the photon-arrival time for a given angular location $\theta$ on the instantaneously-emitting 
 surface, 
\vspace*{-2mm}
\item Equations (\ref{nExpo}) and (\ref{plaw}) are used to determine the Lorentz factor $\Gamma$ at location $\theta(t)$, 
\vspace*{-2mm}
\item Equation (\ref{doppler}) is used to calculate the corresponding Doppler factor, 
\vspace*{-2mm}
\item Equation (\ref{ipEp}) is used to calculate the evolution of spectral characteristics $i'_p(t)$ and $E'_p$, with their normalization 
    during the GRB phase being free fit-parameters (and secondary model parameters),
\vspace*{-2mm}
\item Equation (\ref{lc}) is used to calculate the flux $F_\eps (t)$.
\end{enumerate}

\vspace*{-2mm}
 The purpose of \S\ref{PKLAE} is to allow a test
of the numerical results and an understanding of the differential light-curve features arising for each kind of outflow structure. 

 Both aims are facilitated by the analytical results provided in Table 1 and by the numerical results shown in Figures 1 and 2.
 Figure 1 shows the evolution of $\Gt$, which defines the four GRB and afterglow phases, and of the Doppler factor,
which is the main driver for the afterglow light-curve.
 Figure 2 compares the numerical and analytical asymptotic flux-decay indices, and illustrates that the {\sl Plateaus arising from 
$\n$-Exponential outflows are shorter than those from Power-Law structures, and are followed by faster-falling post-Plateaus}.

 These two conclusions are also illustrated by the $LAE$ model X-ray light-curves shown in Figures 3-6.
 The ability of Power-Law outflows to account better for long Plateaus than $\n$-Exponential structures arises from that
for $\Gctc > 1$, the Plateau dynamical duration $\tpp/\tp \simeq (\Gctc)^{2/(g-1)}$ (Equations \ref{tpPLaw} and \ref{tppPLaw}) becomes 
arbitrarily large as the angular structure exponent $g \ra 1$ from above, while the same dynamical duration for an $\n$-Exponential 
outflow is limited to $\tpp/\tp \simeq 5 (\ln \Gctc)^{2/n}$ (Equations \ref{tpExpo} and \ref{tppExpo}), which is less than a factor 30 
for $\Gctc < 10$ and $n > 1$ (the approximation in Equation \ref{tppExpo} is not good for $n \ra 0$, which would be an outflow with 
a very weak angular structure).

\vspace{2mm}
\subsection{\bf Determination of $LAE$ Model Spectral Slopes}

 For a peak-energy $\Ep$ in the XRT window 0.3-10 keV, the effective slope $\beta$ is between $\bl$ and $\bh$, thus the measured slope
$\beta$ during the GRB prompt phase sets the limits shown in Equation (\ref{blbh}): $\bl \leq \beta (t \ll \to)$ and $\beta (\tp) \leq \bh$. 
 The Core's low-energy spectral slopes $\bl$ for the best fits shown in Figures 3-8 was fixed at a value consistent with that upper limit
$\beta (t \ll \tp)$, but the Core's high-energy slope $\bh$ was left to be a free model parameter, leading to a best-fit spectral slope that, 
more often than not, is harder than the expected limit $\beta (\tp)$. We note that the Core's high-energy slope $\beta_h$ must be a free 
model-parameter because it is primarily determined by the power-law flux-decay index of the Tail ($\alpha_t = 2 + \beta$), which is dependent 
on the best-fit pulse start time $\to$.

 For $\beta (\tp) \leq \bh$, the calculation of Swift fluxes using a high-energy slope $\beta (\sim \tp)$ harder than the slope $\bh$ 
of the best-fit to the already-calculated Swift fluxes is a circular problem that can be solved iteratively to reach values that satisfy 
the "initial" condition $\beta (\tp) \leq \bh$. 

 That recursive calculation would affect mostly the best-fit spectral parameters for the Core emission because our calculation of the Swift 
fluxes employs a broken power-law spectrum only when a clear evolution of the effective slope $\beta$ is seen (indicating a peak-energy 
$\Ep$ in the XRT band), which happens mostly/only during the prompt GRB phase, which arises from the Core. 

 In constrast to the Tail, the general lack of an evolution of the effective slope $\beta$ during the afterglow indicates that the Envelope's 
peak-energy $\Ep$ falls below 0.3 keV, in which case our calculation of the Swift fluxes employs the measured effective slope $\beta$, which 
is consistent with the Envelope high-energy slope $\bh$ because that "consistency" is enforced in the best-fit by requiring $\bh = \beta (t > \tp)$. 

 We note that, also in constrast to the Tail (where the Core's spectral slope $\bh$ is completely {\sl determined} by the Tail flux-decay 
index $\alpha_t$), the best-fit spectral slope $\bh$ of the Envelope emission (which is allowed to be different than the slope $\bh$ of the 
Core emission, to account for the spectral hardening seen at the Tail/Core-Plateau/Envelope transition) is {\sl not constrained} by the 
Plateau and post-Plateau flux-decay indices $\ap$ and $\app$ because these indices also depend on the structural parameter 
$g$ or $n$ and on the spectral parameters $i$ and $e$ for the spectral characteristics peak-intensity $i'_p$ and peak-energy $E'_p$ (Table 1). 
In fact, for $\Ep < 0.3$ keV, the afterglow flux-decay depends on the combination $i + e\bh$, thus these three model parameters are highly
degenerate.

 Unfortunately, for $\bh < \beta (\tp)$, we have reached an incompatibility between the model spectral slope $\bh$ and the effective slope 
$\beta (\tp)$ measured by Swift, which may be due to the assumption that the Core emission has an angle-independent slope $\bh$. 


\vspace{2mm}
\subsection{\bf Constraint on the $LAE$ Model Emission Parameters from Afterglow Light-Curves} 

 The evolutions of the spectral characteristics peak-energy $E'_p \sim \Gamma^e$ and peak-flux $i'_p \sim \Gamma^i$ (Equation \ref{ipEp}) 
determine the (Table 1) only through the combination $i + e\beta$. Consequently, the emission exponents $i$ and $e$ can be determined
individually only when the spectral slopes at 0.3 keV and 10 keV are different (leading to slightly decoupled light-curves), 
i.e. only when the peak-energy $\Ep$ is in the XRT band, which is strongly indicated (although not guaranteed) by an evolving 0.3-10 keV 
effective spectral slope, which is observed predominantly during the Tail. 

 Because the Tail emission is from the outflow Core, the above considerations imply that the emission exponents $i$ and $e$ can be determined 
for a Power-Law outflow with a $\Gamma$ uniform-Core only when one uses an angular $\theta$-paramerization of the spectral characteristics. 
 Thus, with a Lorentz factor $\Gamma$-parameterization of $i'_p$ and $E'_p$, the emission exponents $i$ and $e$ can be determined by the 
Tail emission from the Core only for an $\n$-Exponential outflow. However, such a determination is hampered by the Core's weak angular 
structure, which implies that any differential decay of the 0.3 keV and 10 keV Tail fluxes that is in excess of what is induced by the 
decreasing Doppler factor $\calD \sim t^{-1}$ (and the resulting decreasing peak-energy $\Ep = \calD E'_p$), could require large emission exponents 
$e$ and $i$ because the Tail flux-decay index $\alpha_t$ (Table 1) contains the combination $i + e\beta$ diminished by the sub-unity term
$\onehalf(t/\tp)^{n/2}$. Then, the assumption of constant exponents $i$ and $e$ across the Core-Envelope boundary could easily lead to 
Envelope light-curves whose decays are incompatible with measurements. That the combination $i + e\beta$ cannot be too large to affect
significantly the evolution of $\Ep$ during the Tail (beyond what the Doppler factor $\calD$ can do), is proven by that $\Ep(t)$ 
(shown in lower insets of Figures 3-7) is the same for both the Power-Law outflow best-fit (where $E'_p$ is uniform in the Core) and
the $\n$-Exponential structure (where $E'_p \sim \Gamma^e$ can vary with angle).

 Thus, the outflow angular structures considered here hamper the determination of the secondary emission parameters and the determination 
of the emission parameters $i$ and $e$ from the decoupling of Tail X-ray light-curves requires consideration of stronger (Core) angular 
structures. 

 Afterglow X-ray measurements correspond to a peak-energy $\Ep$ that is most likely below the XRT band, as indicated by the constancy 
of the effective spectral slope $\beta$. Thus, lacking a decoupling of the 0.3 keV and 10 keV light-curves, afterglow X-ray measurements 
can only constrain the combination $i + e\bh$, with $\bh$ the high-energy slope of the Envelope emission (which is just the effective slope 
$\beta$ measured by XRT).
 One can use the asymptotic Plateau and post-Plateau flux-decay indices $\ap (\tp)$ and $\app (\tpp < t)$ 
for an $\n$-Exponential outflow, and $\ap (\tp < t < \tpp)$ and $\app (\tpp < t)$ for a Power-Law outflow, given in Table 1, to show that 
\begin{equation}
 \hh {\rm (n-Exp)} \quad \app - \ap(\tp) = 1 ,\; i + e\beta = 2(\app-2) - \beta
\label{ieExp}
\end{equation}
for an $\n$-Exponentail outflow and
\begin{equation}
 \hspace*{-4cm} {\rm (P-Law)} \quad g = 1 + \frac{\app - \ap}{\beta + 3 - \app} 
\label{gPL}
\end{equation}
\begin{equation}
 i + e\beta = 2(\app-1) - \beta - \frac{2(\app - \ap)}{\beta + 3 - \ap}
\label{iePL}
\end{equation}

 Equation (\ref{ieExp}) relates the Plateau and post-Plateau asymptotic flux-decay indices, and provides a possible 
{\sl test for the $\n$-Exponential structure}, with the caveat that the flux-decay index $\ap(\tp)$ may be difficult to measure 
accurately from X-ray light-curves.

 Equation (\ref{gPL}) provides a new determination of the structural index $g$ for a Power-Law outflow, which, when added to its 
determination from the afterglow temporal breaks (Equation \ref{ng}), leads to a potential {\sl test for the Power-Law structure},
with the caveat that it may be dificult to measure the index $g$ using the afterglow temporal milestones.

 Finally, Equations (\ref{ieExp}) and (\ref{iePL}) provide the only {\sl constraint} that {\sl afterglow} X-ray observations 
can set on the emission parameters. For the majority of the Power-Law structure best-fits shown in Figures 3-8, the combination 
$i+e\bh \in [-1.5,-1]$ and the high-energy spectral slope of the Envelope emission $\bh \simeq 1$, leading to that the emission 
parameters satisfy 
\begin{equation}
 (i+e)_{aglow} \in [-1.5,-1]
\label{ieaglow}
\end{equation}

 {\sl If} the spectral characteristics dependences given in Equation (\ref{ipEp}) depend only on the Lorentz factor $\Gamma$, 
i.e. the normalization constants and the indices $e$ and $i$ are the same for all bursts (thus, this is a very big "if"), 
then $i'_p \sim \Gamma_c^e$ and $E'_p \sim \Gamma_c^i$ for the prompt phase ($t < \tg$) of all bursts. For the $LAE$ model, 
the corresponding observables satisfy 
$\Ep = \calD E'_p \sim \Gamma_c^{e+1}$ (peak-energy), $F_p \sim \calD^2 i'_p \sim \Gamma_c^{i+2}$ (peak-flux), and the bolometric
GRB pulse fluence is $\Phi \sim \Ep F_p \tg \sim \Gamma_c^{i+e+3} \to \sim \Gamma_c^{i+e+1} t_{lab}$, after using $\tg \simeq 2\to$ (Table 1)
and Equation (\ref{to}). If a last bold assumption, that the GRB emission is produced at a universal location $ct_{lab}$ that 
is independent of $\Gamma_c$, is made, that would lead to $\Ep^{i+e+1} \sim \Phi^{e+1}$ for all GRBs. 

 Lloyd, Petrosian and Mallozzi (2000) found that, for a small sample of bright BATSE GRBs, the burst-integrated fluence $\Phi_\gamma$
at 25 keV -- 1 MeV is well-correlated with the peak-energy: $\Ep \sim \Phi_\gamma^{1/2}$, with that correlation not being due to cosmological 
effects, i.e. is not caused by the spread in redshift of those bursts. Also for a small set of bright bursts, Amati et al (2002) have shown 
that the intrinsic peak-energy $(z+1)\Ep$ and GRB isotropic-equivalent output at $1-10^4$ keV ${\cal E}_\gamma \sim \Phi_\gamma D_L^2/(z+1)$ 
are well-correlated, with a similar dependence $(z+1)\Ep \sim {\cal E}_\gamma^{1/2}$. 

 Put together with the above correlation established for the $LAE$ emission, this peak-energy--isotropic-output correlation, measured
for real GRBs, implies that 
\begin{equation}
  (i+e+1)_{grb} = 2\, (e+1)_{grb} \; \lra \; (i-e)_{grb} = 1 
\end{equation}
which, together with the Equation (\ref{ieaglow}) for the $LAE$ model to account for the afterglows of Figures 3-8, leads to 
$e = \onehalf[(i+e)_{aglow}-1]$ and $i = \onehalf[(i+e)_{aglow}+1]$, with $(i+e)_{aglow}$ a universal value that satisfies Equation (\ref{ieaglow}). 

 For $(i+e)_{aglow} = -1$, one obtains $e=-1$ and $i=0$, while $(i+e)_{aglow} = -1.5$ leads to $e=-1.25$ and $i=-0.25$.
 Thus, if the $LAE$ spectral characteristics depend only on the Lorentz factor $\Gamma$, and the indices $e$ and $i$ and the GRB source 
location were all universal, then the $\Ep \sim {\cal E}_\gamma^{1/2}$ correlation measured for real bursts implies that the GRB mechanism 
is one that yields a comoving-frame peak-energy that increases with decreasing Lorentz factor (which is opposite to that expected for the 
synchrotron and inverse-Compton emissions from the forward-shock driven by the relativistic ejecta into the ambient medium) and a comoving-frame 
peak-intensity that is weakly-dependent on $\Gamma$ in the outflow Envelope.

 Nevertheless, such conclusions are unreliable because of the underlying assumption that the comoving-frame spectral characteristics
depend only on the local Lorentz factor. In real bursts, those spectral characteristics may also depend on other properties of the GRB ejecta,
such as their kinetic energy density or the local magnetic field, and, if the angular distributions of those quantities are not correlated
with the $\Gamma (\theta)$ angular structure, then the underlying assumption becomes incorrect.

\vspace{2mm}
\subsection{\bf Numerical Fits to GRB X-ray Afterglows}
\label{NumFits}

\subsubsection{Numerical Model Parameters} 

 The numerical $LAE$ model has all the "analytical" parameters presented in \S\ref{params}: 
two {\sl primary structural} parameters, 
two {\sl secondary emission} parameters, 
and three {\sl tertiary spectral} parameters. 

 Additionally, the numerical $LAE$ model has two {\sl temporal} parameters: the epochs $\To$ when the first photons arrive and $\Tg$ when 
the GRB prompt phase ends.
 These two {\sl epochs}, indicated in Figures 3-8 relative to the trigger time, are related to the corresponding {\sl durations} $\to$ 
and $\tg$ given in Equations (\ref{to}) and (\ref{tgrb}). 
 The numerical parameter $\To$ is left free to minimize the $\chi^2$ contribution arising from the "instantaneous emission" assumption of 
the $LAE$ model, which introduces a discrepancy between a real GRB pulse shape, displaying a slow rise, and the $LAE$ model pulse shape 
that exhibits an infinitely sharp rise.
 The numerical parameter $\Tg$ accounts for the GRB pulse duration, which is equal to $\Tg - \To$, and which should be close to 
$\tg-\to \simg \to$ (Equation \ref{tgrb}), with $\to$ defined in Equation (\ref{to}). Thus, $\Tg$ is a normalization parameter.

 Finally, the numerical $LAE$ model has two parameters for {\sl spectrum normalization}: the peak-energy $\Ep$ and peak-flux $F_p$ at the 
beginning of the GRB pulse.

\subsubsection{Best-Fits by $\chi^2$-Minimization}

 Best-fits are determined by $\chi^2$ minimization between the XRT 0.3 keV and 10 keV flux densities (calculated
as described above) and the $LAE$ model flux (calculated also as described above). For clarity, in Figures 3-8, measurements have been rebinned 
within a 2-3\% time-interval, and the averages and errors are error-weighted, thus the $\chi^2$ statistics should be the same, whether binned
or unbinned data are used. 
  
 For some afterglows, the best-fits accommodate well all light-curve features. Adding that our calculation of flux errors takes into account 
both the measurement uncertainties of the blometric flux and of the effective slope, strongly indicates that the improbable, above-unity 
$\chi^2_\nu$ of those better best-fits should be blamed on short-term light-curve fluctuations that cannot be accounted for by the $LAE$ 
from an outflow with a smooth, monotonic angular structure.

 Light-curve discontinuities for the emission from the Core-Envelope boundary ($\theta(\tp) = \theta_c$) are due to the discontinuous
high-energy slope $\bh$ at that boundary, which is required by the hardening (sudden or gradual) of the XRT effective slope $\beta$ 
across $\tp$.
 For the Power-Law angular structure model, light-curves are also discontinuous because the observed $LAE$ flux is proportional to 
$d\theta/dt$, which is itself discontinuous at the Core-Envelope boundary.

\subsubsection{Spectral Hardening at Tail End}

 Some best-fits fail to accommodate the Tail end and Plateau beginning of those afterglows that display decoupled 0.3 keV and 10 keV
light-curves at those phases. That decoupling arises from an unusual spectral hardening at the Tail end.
 While a Tail softening is a natural feature of the $LAE$ model (upper panel of Figure 2), where a spectral softening is due to a decreasing 
peak-energy $\Ep$ (upper panel of Figure 2), which is due to a decreasing Doppler factor $\calD$ (lower panel of Figure 1), 
a tail hardening is not a natural feature of the $LAE$ model as considered here, and requires either an increasing peak-energy $\Ep$ for the 
Core emission or a hardening of the high-energy slope $\bh$ of the Core X-ray continuum.

 For the former case, if $\Ep$ were to increase at the Tail end and stay in the XRT band to account for $\beta \in [\bl,\bh]$ for the Core
emission, then the transition to a constant spectral slope $\beta$ for the afterglow Envelope emission implies that the peak-energy
$\Ep = {\calD} E'_p$ should remain in the XRT band for the entire afterglow, but that feature cannot be accommodated by the $LAE$ model because, 
in it, the Doppler factor evolution changes from nearly constant during the Plateau to a decreasing $t^{-1/2}$ during the post-Plateau (Table 1).

 Thus the spectral hardening of the Tail end is more likely due to a hardening of the high-energy slope $\bh$ of the Core emission, 
a feature that is not included in our $LAE$ model.  
 Instead, a high-energy slope $\bh$ that depends on the angular location (or on the Lorentz factor) is allowed only as a sharp transition 
at the Plateau beginning ($\tp$) from a constant value in the Core to a constant value in the Envelope, but that is not flexible enough 
and leads to poor fits at the Tail end.

 However, the above deficiency has no effect on assessing the ability of angular structures to accommodate the afterglow emission arising 
from the outflow Envelope. As discussed before, a Power-Law outflow has a natural advantage over an $\n$-Exponential one in producing Plateaus 
of a larger dynamical range $\tpp/\tp$, whose value has a {\sl much stronger dependence on the structural index $g$ of a Power-Law outflow
than on the index $n$ of a generalized-Exponential structure} (Table 1). 
 Additionally, Power-Law outflows perform better than $\n$-Exponentials because they yield post-Plateaus flux decays that are slower
by $\delta \app = 1/g \simeq 1/2$ (Table 1). 

 From a numerical point of view, these two limitations of $\n$-Exponential outflows lead together to that, {\sl for a smaller $\Gctc$, 
this structure can account for the Plateau flux but could under-predict the post-Plateau while, for a larger Core parameter $\Gctc$, 
it can account for the post-Plateau emission but could underpredict the Plateau flux}.

\subsubsection{Calculation of Flux-Densities from XRT Data}

 The upper insets in the upper panels of Figures 3-7 show the ratio of fluxes calculated from the XRT data using a broken power-law spectrum 
(as described above)
to those corresponding to a pure power-law spectrum of effective slope $\beta$ (Equation \ref{flux}). That ratio is below unity only when the 
peak-energy $\Ep$ of the broken power-law spectrum is in the XRT 0.3-10 keV window, i.e. when there is a spectral evolution of the XRT effective 
slope $\beta$ in the lower panel, which is observed only during the Tail. Because no significant/systematic spectral evolution is measured during
the afterglow, we assumed that $\Ep < 0.3$ keV after the Tail end at $\tp$, and the calculation of the 0.3 keV and 10 keV fluxes employs a pure 
power-law spectrum with the XRT effective slope $\beta$. 

 We note that, during the Tail, the ratio of fluxes calculated using a broken power-law to those obtained using a single power-law is 
energy-dependent and is a factor that decreases for a softer spectral slope at the observing energy. Both these findings illustrate 
{\sl the importance of using a broken power-law spectrum to calculate flux densities from XRT data when a spectral evolution is present}.

\subsubsection{XRT and $LAE$-Model Peak-Energies $\Ep$}

 The lower insets in the upper panels of Figures 3-7 show the peak-energy $\Ep$ of the broken power-law spectrum calculated for the XRT data 
(symbols with error bars) using Equations (\ref{Eplo}), (\ref{Ephi}), and (\ref{sigEp}), and for the indicated spectral slope $\bl$ and $\bh$, 
chosen according to Equation (\ref{blbh}), compared with direct measurements of that peak-energy from fits to the XRT continuum, whenever
they were available. 

 The consistency shown in Figures 4 and 5 between our speedy "measurement" of peak-energy $\Ep$ and the spectral-fit measurements indicates 
that the high-to-low {\sl count}-ratio $\HR_c (\beta) \equiv C(1.5-10)keV/C(0.3-1.5)keV$ reported by XRT is a good proxy for the {\sl photon-flux} 
ratio $\HR (\beta)$ in Equation (\ref{HR}).

 The lower insets also shows the evolution of the peak-energy $\Ep$ of the best-fit $LAE$. 
In general, consistency between the $LAE$ model peak-energy $\Ep$ and the values inferred from XRT measurements is only approximative/tentative 
for the Core/GRB emission, with the former underestimating the latter by 
a factor up to two. We note that model best-fit are obtained by matching the fluxes "measured" (rather "inferred" from measurements) at 0.3 keV 
and 10 keV, and does not include matching peak-energies, whose "measured" values depend on the choice of the asymptotic spectral slopes $\bl$ 
and $\bh$. In fact, Figures 3-7 show that this peak-energy mismatch is correlated with the high-energy spectral slope $\bh = \beta(\tp)$ 
being softer than the $LAE$ model slope $\bh = \alpha_t - 2$ (shown in lower panels), which is set by the Tail flux-decay $\alpha_t$.

\subsubsection{XRT and $LAE$-Model Effective Spectral Slopes $\beta$}

 The lower panels of Figures 3-8 show the XRT effective spectral slope $\beta$ calculated by the Swift team, and represents the slope
of a single power-law spectrum that would account for the hardness-ratio $\HR_c$ that was measured by XRT.
 Lines show the effective slope $\beta$ for the $LAE$ best-fit model, i.e. the slope of a single power-law spectrum that would yield 
a photon-flux ratio $\HR$ (Equation \ref{HRbeta}) equal to that resulting for the $LAE$ best-fit.
 
 A good agreement between the effective slopes $\beta$ measured by XRT for the Core/GRB emission and those resulting for the $LAE$ model 
best-fits is quite elusive, but that is not due our equating photon-flux ratios $\HR$ with count-rate ratios $\HR_c$, because that confounding 
did not introduce significant errors in estimating the peak-energy energy $\Ep$.

 For peak-energies, we note that the data-derived effective slope $\beta(t)$ is not used for finding the best-fits because all the 
information contained in that slope has been transferred to the 0.3 keV and 10 keV light-curves. Furthermore, the mismatch between measurements 
and $LAE$ model best-fit results can also be attributed to the mismatch between the asymptotic low and high-energy slopes $\bl$ and $\bh$
implied by XRT measurements and best-fit values that are determined by the Tail flux decay, owing to that $\alpha_t = \beta + 2$: 
for 0.3 keV$< \Ep <$ 10 keV (Tail emission from the Core), $\bl$ ($\bh$) is set by the decay of the 0.3 (10) keV flux.

\vspace{2mm}
\section{\bf Discussion}

 The formalism developed here for the calculation of $LAE$ light-curves from the photon-arrival time and Doppler boost of the 
comoving-frame emission and for two angular structures -- {\sl a generalized Exponential and a Power-Law} -- can be used to identify
light-curve features specific to each structure ($\n$-Exponential outflows yield shorter afterglow Plateaus, followed by steeper 
afterglow post-Plateau decays, than Power-Law structures) and to fit XRT light-curves, with the purpose of {\sl identifying 
the angular structure} that accommodates observations the best. 

 In addition to the outflow angular structure, afterglow light-curves are also determined by how the {\sl spectral characteristics} of 
the {\sl comoving-frame} broken power-law spectrum -- peak-energy $E'_p \sim \Gamma^e$ and peak-intensity $i'_p \sim \Gamma^i$ -- change 
across the emitting surface. However, these secondary factors cannot erase or mimic the features imprinted by the angular structure, 
as long as they have "reasonable" distributions on that surface.

 Through fits to light-curves of six XRT light-curves of all existing types (with or without an afterglow Plateau, with or without 
a GRB Tail), we found that, indeed, $\n$-Exponential outflows can accommodate only afterglows with a Plateau dynamical range shorter 
than about 1.5 dex (Figures 3-4), that the following post-Plateau steeper flux-decay cannot be mitigated by an adequate evolution of 
the spectral characteristics (or, rather, an adequate $i + e\beta$ combination, with $\beta$ the X-ray spectral slope) without compromising
the model good-fit to the GRB Tail emission, leading to steep post-Plateau decays that are incompatible with the observations for Plateaus 
lasting for more than 2 dex (Figures 6-7) or for afterglows without Plateaus (Figure 8).

 A determination of the emission exponents $i$ and $e$ can be done only when the peak-energy $\Ep$ is in the XRT 0.3-10 keV band,
i.e. when the 0.3 keV and 10 keV light-curves display some {\sl decoupling} (or when the X-ray continuum shows a spectral evolution) in excess 
of what the natural evolution of the observer-frame peak-energy $\Ep = \calD E'_p$ (caused by the evolution of the relativistic Doppler 
factor $\calD$ across the emitting surface) would induce. XRT observations show that a clear spectral evolution occurs only (or mostly) 
during the GRB Tail, and not during the afterglow phase, which implies that:
 $i)$ $\Ep$ is in the 0.3-10 keV band only for the outflow Core's emission and that would allow the {\sl determination} of the spectral 
characteristics evolution, provided that a strong(er) angular structure is considered, and
 $ii)$ $\Ep$ is below the 0.3-10 keV band for the outflow Envelope's emission and that provides only a {\sl constraint} on the emission 
exponents $i$ and $e$, with a complete determination possible if optical observations (below $\Ep$) are also included and attributed
to the same $LAE$ (i.e. for well-coupled optical and X-ray light-curves).

 Because the above-discussed determination of spectral characteristics evolution relies on modeling the decoupling of X-ray light-curves 
during a period of spectral evolution, we have developed a calculation of the 0.3 keV and 10 keV flux-density histories from the XRT 
0.3-10 keV bolometric fluxes and effective spectral slopes $\beta$ that is based on determining first the peak-energy $\Ep$ of a broken 
power-law spectrum of constant low-energy (below $\Ep$) and high-energy (above $\Ep$) slopes ($\bl$ and $\bh$) that are set 
by the asymptotic measured effective slope $\beta$ at the GRB pulse beginning and Tail end, respectively: $\bl = \beta (t \ll \tg)$
and $\bh = \beta (\tp)$. That $\Ep$ determination is based on requiring that the broken power-law spectrum yields an effective slope
$\beta$ equal to that inferred by the XRT team from the measured high-to-low-energy (hardness) count-ratio.

 This method for a fast determination of $\Ep$ from XRT data fares well when comparing it with $\Ep$ measurements obtained through more 
"expensive" direct fits with a broken power-law spectrum to the XRT continuum (which are limited to only the brighter bursts). 
While a decoupling of the 0.3 keV and 10 keV X-ray light-curves can be extracted from the XRT data by employing a pure power-law spectrum, 
the exact decoupling (useful for a correct determination of model features) of those light-curves requires the use of a broken power-law 
spectrum. We note that the $LAE$ model provides {\sl always} a better fit to the decoupled 0.3 keV and 10 keV Tail light-curves 
(i.e. when the peak-energy $\Ep$ crosses the XRT band) calculated with a broken power-law spectrum than to the decoupled Tail light-curves
calculated with a pure power-law XRT continuum. For the GRB Tails shown in Figures 3-7, the corresponding increase in $\chi^2_\nu$ ranges 
from 5\% to 30\% and is statistically significant, even though all best-fits are statistically unacceptable. 

 This best-fit improvement is not a proof that the $LAE$ model for GRBs and X-ray afterglows is correct (that is proven by this model's 
natural ability to account for the X-ray afterglow diversity), nor is it evidence that the calculation of the 0.3 keV and 10 keV Tail fluxes 
should be done using a broken power-law spectrum (that is motivated by the spectral softening of GRB Tails, which is more naturally 
attributed to a decreasing peak-energy $\Ep$ of a broken power-law spectrum than to a softening of an otherwise power-law X-ray continuum). 
Instead, the best-fit improvement just makes the case that a broken power-law spectrum should be used for a more accurate determination 
of the $LAE$ model best-fit (structural and emission) parameters.

\vspace{3mm}
\section{\bf Conclusions}

 This work's purpose {\bf is not to test the $LAE$ model} because its applicability to X-ray afterglows is proven by its ability to account 
so simply for afterglow Plateaus and post-Plateaus. 

 Instead, the {\bf first aim} is to show that {\sl XRT observations can constrain the outflow angular structure}, in the hope that determinations 
of the Lorentz factor distribution $\Gamma (\theta)$ will serve as {\sl benchmarks for simulations of relativistic jets} produced by 
solar-mass black-holes, propagating through the infalling envelopes of the massive stars that undergo core-collapse and through 
the supernova ejecta driven by such objects, while recognizing that it may be very difficult to disentangle the features imprinted 
on the outflow angular structure by the jet production (determined by the progenitor type, its spin and magnetic field geometry/structure, 
mass and angular momentum of the accretion disk) from those imprinted by the jet propagation through the progenitor star
(set by the jet initial opening angle, density of the ambient medium, jet recollimation by the cocoon formed after the jet lateral 
expansion). 
 
 For this aim, we find that outflows endowed with a power-law angular structure (such as that obtained by Urrutia, De Colle \& Lopez-Camara 2023
in hydrodynamic simulations of jets propagating through their progenitors and the SN ejecta) are outperforming $\n$-Exponential outflows 
(such as the exponential and Gaussian profiles identified by McKinney 2006 from MHD simulations of Poynting jets). 
More complex angular structures, such as the three power-law $\Gamma(\theta) \sim \theta^{-g}$ with $g=-1,1,2.5$ 
identified by Tchekhovskoy A. et al (2008) through simulations of magnetodynamic jets, lead to X-ray power-law light-curves whose decay
indices $\alpha_{p,pp}$ (Table 1) are incompatible with observations.

 The {\bf second aim} is to assess if the spectral characteristics peak-intensity $i'_p \sim \Gamma^i$ and peak-energy $E'_p \sim \Gamma^e$ 
dependences on the Lorentz factor $\Gamma$ can be constrained using the XRT data. 

 A determination of the indices $e$ and $i$ can be done only if the peak-energy $E_p$ crosses the 0.3-10 keV range during the Plateau (or post-Plateau), 
to break-up the light-curve decay indices $\alpha_p$ and $\alpha_{pp}$ degeneracy on the emission angular indices $i$ and $e$  (Table 1). 
The $E_p$ crossing the observations window entails a spectral softening of the X-ray continuum. Measurements of the peak-energy $E_p$
for the X-ray afterglows modeled here suggest that $E_p$ falls below the XRT window before or around the Plateau beginning, which is consistent
with a lack of spectral softening of the XRT spectrum during the Plateau. Thus, given that the latter is a feature of most XRT X-ray Plateaus, 
X-ray measurements alone can only constrain the combination $i+e \beta$ and cannot determine each index separately. For that, {\sl optical measurements}
(and an afterglow with well-coupled optical and X-ray light-curves) are needed to break the degeneracy of the flux power-law decay indices 
$\alpha_{p,pp}(i,e)$.

 The best-fits with the $LAE$ model to six X-ray afterglows (Figures 3-8) satisfy $i+e \beta < 0$. Given that the XRT spectral slope is $\beta \simeq 1$,
this constraint implies that one or both spectral peak characteristics $i'_p$ and $E'_p$ increase with decreasing Lorentz factor in the outflow Envelope. 
That seems at odds with a shock origin of the prompt GRB emission, where one might expect a higher electron energy and magnetic field (leading to a 
larger $i'_p$ and $E'_p$) for a higher outflow Lorentz factor.

 If the spectral characteristics $i'_p$ and $E'_p$ depend only on the local Lorentz factor, 
then constraining or determining the emission characteristics exponents $e$ and $i$ could lead to a {\sl test for the GRB dissipation mechanism} 
(magnetic reconnection, internal shocks, external shock) and for the {\sl GRB emission process} (synchrotron and inverse-Compton).
 For instance, if the spectral characteristics have universal coefficients and exponents in Equation (\ref{ipEp}), then adding the peak-energy
$\Ep$ -- isotropic-equivalent output correlation $\Ep \sim {\cal E}_\gamma^{1/2}$ measured for long GRBs (Lloyd et al 2000, Amati et al 2002),
leads to $e \simeq -1$ and $i \simeq 0$.
 The former inequality rules out the forward-shock as the GRB dissipation mechanism.
 If the spectral characteristics $i'_p$ and $E'_p$ depend significantly on other quantities, such as the ejecta kinetic-energy per solid angle
or their magnetization, then identifying the GRB dissipation mechanism (and emission process) becomes more complicated or, maybe, impossible.
 
 We conclude by proposing that modeling Swift/XRT afterglows with the $LAE$ model offers the best avenue to measure the {\bf angular} 
structure of GRB outflows, especially if the optical emission is included whenever that light-curve is {\bf well-coupled} (achromatic breaks 
and similar decays) with the X-ray's. Through its two angular structureal parameters, the $LAE$ model can account easily for the observed
diversity of XRT light-curves, from those with a GRB tail and an afterglow plateau to those without either.

 The afterglow emission at lower energies (optical and radio), produced by the relativistic blast-wave driven by the GRB ejecta into the ambient 
medium, is set not only by the outflow angular structure but also by its {\bf radial} structure, which leads to energy-injection in the blast-wave. 
 Economy of assumptions indicates that the external-shock emission should be the dominant mechanism only for the sub X-ray emission of afterglows 
with decoupled X-ray and optical light-curves (chromatic X-ray light-curve breaks, different flux decays).


\acknowledgments{ {\sl Acknowledgments:}
  This work made use of data supplied by the UK Swift Science Data Centre at the University of Leicester:
  {\sl www.swift.ac.uk/burst\_analyser} (Evans et al 2010) }



\appendix

\section{\bf A. How to Calculate X-ray Flux Densities from the XRT Data}

 An unevolving effective slope $\beta$ strongly suggests that peak-energy $\Ep$ of the synchrotron spectrum is not in the XRT band
because the opposite of that  
requires a conspiracy between the Doppler factor $\calD$ and the comoving-frame peak-energy $E'_p$ to yield a constant $\Ep = \calD E'_p$.
 In this case, $\beta$ represents an measurement of the real slope of the 0.3-10 keV continuum: $F_\eps \sim \eps^{-\beta}$. 
Then, the flux density at any energy $\eps \in [0.3,10]$ keV is
\begin{equation}
 F_\eps = \frac{\ds 1-\beta}{\ds \left[ \left( \frac{10\k}{\eps} \right)^{1-\beta} \hh - \left( \frac{0.3\k}{\eps} \right)^{1-\beta} \right]} 
    \frac{\ds \F0310k}{\ds \eps}
\label{flux}
\end{equation}

 An evolving effective slope $\beta$ at 0.3-10 keV is compatible with either the peak-energy $\Ep$ being in that band or with it being below
it and with an evolving high-energy slope $\bh$. After dismissing the latter possibility by making the assumption that the spectral slope 
$\bh$ is angle-independent, except for a discontinuous change at the Core-Envelope boundary (corresponding to the observed spectral hardening 
at the Tail-Plateau transition), we note that the use of a variable effective slope $\beta$ in the above equation would lead to erroneous flux 
densities simply because the real spectrum is not a power-law but a broken power-law. Then, the error made by using Equation (\ref{flux}) is 
a factor between 0.1 and 5 that is energy-dependent, thus the use of this equation will introduce an otherwise non-existing decoupling of 
the 0.3 keV and 10 keV light-curves.

 To avoid introducing such a challenging and artificial feature, one should calculate the flux densities using a broken power-law spectrum
(Equation \ref{spectrum})
of low and high-energy slopes $\bl$ and $\bh$ compatible with (or set by) the effective slope $\beta$ at the beginning and end, respectively, 
of the phase where an evolving slope $\beta$ is measured. For the GRB prompt phase, that means that 
\begin{equation}
 \bl = \beta (t \ll \to) \;,\quad \bh = \beta (\tp)
\label{blbh}
\end{equation}

 The remaining issue is to calculate the peak-energy $\Ep$ of that broken power-law
spectrum from the measured evolving effective slope $\beta$. That is done by requiring that a broken power-law spectrum of (known) low and
high-energy slopes $\bl$ \& $\bh$ and peak-energy $\Ep$ yields a high-to-low-energy photon-flux ratio 
\begin{equation}
 \HR \equiv \frac{F(1.5-10)keV}{F(0.3-1.5)keV}
\end{equation}
equal to that resulting if the input spectrum were a pure power-law with the effective slope $\beta$ reported by XRT.

 Thus, we solve numerically for $\Ep (\beta)$ in 
\begin{equation}
  \HR (\beta) = \HR (\Ep) 
\label{HR}
\end{equation}
where 
\begin{equation}
  \HR (\beta) = \frac{\ds 10^{-\beta}-1.5^{-\beta}} {\ds 1.5^{-\beta}-0.3^{-\beta} } 
\label{HRbeta}
\end{equation}
\begin{equation}
 \HR (\Ep) = \frac{\ds \frac{\bl}{\bh} \left[1 - \left( \frac{\Ep}{10\k}\right)^{\bh} \right] + \left( \frac{\Ep}{1.5k} \right)^{\bl} - 1 } 
    {\ds \left( \frac{\Ep}{0.3\k} \right)^{\bl} - \left( \frac{\Ep}{1.5k} \right)^{\bl} }  
\label{Ephi}
\end{equation}
for $\Ep \in [1.5,10]$ keV and
\begin{equation}
 \HR (\Ep) = \frac {\ds \left( \frac{\Ep}{1.5k} \right)^{\bh} - \left( \frac{\Ep}{10\k} \right)^{\bh} }  
    {\ds \frac{\bh}{\bl} \left[ \left( \frac{\Ep}{0.3\k}\right)^{\bl} - 1 \right] - \left( \frac{\Ep}{1.5k} \right)^{\bh} + 1 } 
\label{Eplo}
\end{equation}
for $\Ep \in [0.3,1.5]$ keV.

 From here, the peak-flux $F_p$ for a broken power-law spectrum, at the peak-energy $\Ep$, can be calculated: 
\begin{equation}
 F_p = \frac{\ds \F0310k}{\ds \Ep} 
 \hh \left\{ \h \frac{\ds 1}{\ds 1\h -\h \bl} \h \left[ 1 \h - \h \left( \frac{0.3\k}{\Ep} \right)^{\h 1-\bl} \h \right] \h + 
      \frac{\ds 1}{\ds 1\h -\h \bh} \h \left[ \left( \frac{10\k}{\Ep} \right)^{\h 1-\bh} \h \hh - \h 1 \right] \right\}^{\h -1} 
\label{Fp}
\end{equation}
and the uncertainty in the peak-energy $\Ep$, corresponding to an error $\sigma(\beta)$ in the effective slope, is calculated numerically 
by differentiation of Equation (\ref{HR}):
\begin{equation}
  \sigma(\Ep) = \frac{|d\HR(\beta)/d\beta|}{|d\HR(\Ep)/d\Ep|} \sigma(\beta)
\label{sigEp}
\end{equation}
 
 The 0.3 keV and 10 keV light-curves shown in the upper panels of Figures 3-8 are calculated from the XRT bolometric flux $\F0310k$ 
and effective slope $\beta$ as discussed above, using a {\sl pure power-law spectrum} of slope $\beta$ in Equation (\ref{flux}) 
when that slope is not evolving (i.e. {\sl for the afterglow}), and a {\sl broken power-law spectrum} with the low- and high-energy slopes 
$\bl$ and $\bh$ given in Equation (\ref{blbh}) and of peak-energy $\Ep$ and peak-flux $F_p$ determined from Equations (\ref{Ephi})-(\ref{Eplo})
and Equation (\ref{Fp}) when that slope is evolving (i.e. {\sl for the GRB phase}).
 To avoid a double accounting of the slope uncertainty $\sigma(\beta)$, the flux-density errors are calculated using Equation (\ref{flux})
even for the GRB phase, when the spectrum is a broken power-law.

\vspace*{-1cm}
\begin{table*}
\caption{\small \vspace*{2mm} 
  {\bf Summary of important analytical results}:
  asymptotic solutions $\theta(t)$ to the photon-kinematic equation (Equation \ref{pke}) for the one-to-one correspondence between angular 
  location $\theta$ on the emitting surface and observer photon-arrival time $t$ (first photons arrive at $t(\theta=0)=\to$), 
  evolution of Lorentz factor $\Gamma$ and Doppler factor $\calD$ (Figure 1), 
  afterglow temporal milestones ($\tg$, $\tp$, $\tpp$), 
  and indices $\alpha$ (Equation \ref{alpha}) for the power-law flux (Equation \ref{LAEflux}) decay, 
  for an outflow endowed with an $\n$-Exponential and a $g$-Power-Law angular structure.
  $\beta$ is the spectral slope, 
  $i$ and $e$ are the exponents for the evolution of the comoving-frame spectral characteristics (Equation \ref{ipEp}). 
  The index $\alpha$ is shown in Figure 2 for $i=e=0$.
  Because all relevant quantities $\theta$, $\Gamma$, $\calD$, and $\alpha$ are asymptotic, continuity at the temporal milestones 
  should not be expected.
  Phases 2 (Tail) and 3 (Plateau) do not exist for a generalized-Exponential outflow if $\Gctc < e^{1/n}$, 
  or for a Power-Law outflow if $\Gctc < 1$ and $g > 1$.
  For a Power-Law outflow with index $g < 1$, Phase 1 (GRB) is followed by Phase 4 (but with an increasing $\Gt < 1$) and then Phase 3 
  (but with an increasing $\Gt > 1$) if $\Gctc < 1$, and is followed by Phase 2 (Tail) and then by Phase 3 (but with an increasing $\Gt > 1$)
  if $\Gctc < 1$, thus $g < 1$ yields post-GRB phases (from the Envelope emission) of ambiguous designation.
  {\bf $\n$-Exponential outflows yield two or four GRB \& afterglow phases, while Power-Law outflows yield two, three, or four such phases}.
  }
\vspace*{5mm}
\centerline{
\begin{tabular}{r|l}
  \hline \hline \\
 GENERALIZED-EXPONENTIAL $\Gamma = \Gamma_c e^{-(1/n)(\theta/\theta_c)^n}$ ($n>0$) & POWER-LAW $\Gamma = \Gamma_c (\theta/\theta_c)^{-g}$ ($g>1$) \\ 
  \\ \hline \hline  \multicolumn{2}{c}{} \\
 \multicolumn{2}{c}{1. GRB from CORE ($\theta < \theta_c$), $\to < t < \tg$, increasing $\Gt < 1$ ({\bf angular-structure} term dominant in Eq \ref{pke})} \\ 
       \multicolumn{2}{c}{} \\ \hline \\
 ($n < 2$) : $\theta \simeq \theta_c \left[\frac{n}{2}(t/\to - 1)\right]^{1/n}$, 
             $\Gamma = \Gamma_c e^{-(1/2)(t/\to - 1)}$                &  (uniform Core)  \\
 ($n > 2$) : $\theta \simeq \Gamma_c^{-1} (t/\to - 1)^{1/2}$          &  $\theta$ same as for $\n$-Exponential (left)   \\
 ($n > 2$) : $\Gamma = \Gamma_c e^{-(1/n)[(t-\to)/\tp]^{n/2}}$        &  $\Gamma = \Gamma_c$ \\
             $\calD \simeq 2\Gamma$, $\alpha_\gamma \simg 0$          &  $\calD \simeq 2\Gamma_c$, $\alpha_\gamma = 0$  \\
             $\tg \simeq 2\to$                                        &  $\tg$ same as for $\n$-Exponential (left)    \\
       \\ \hline \multicolumn{2}{c}{} \\
 \multicolumn{2}{c}{2. TAIL from CORE, $\tg < t < \tp$, increasing $\Gt > 1$ ({\bf spherical-curvature} term dominant} \\ 
         \multicolumn{2}{c}{} \\ \hline \\
   $\theta = \theta_c [(t-\to)/\tp]^{1/2}$                            &  $\theta$ same as for $\n$-Exponential (left)   \\
   $\Gamma$ same as for GRB for $n > 2$ (above)                       &  $\Gamma = \Gamma_c$ \\
   $\calD \sim t^{-1} e^{(1/n)(t/\tp)^{n/2}}$                         &  $\calD \sim t^{-1}$ \\
   $\alpha_t = (2+\beta) + \onehalf (i+e\beta -2-\beta) (t/\tp)^{n/2}$ ($t \ll \tp: \alpha_t \simeq 2 + \beta$)  &  $\alpha_t = 2 + \beta$ \\
   $\tp = [(\Gctc)^2 + e^{2/n}] \to$                                  &  $\tp = [(\Gctc)^2 + 1] \to$ \\
       \\ \hline \multicolumn{2}{c}{} \\
 \multicolumn{2}{c}{3. PLATEAU from ENVELOPE ($\theta_c < \theta$), $\tp < t < \tpp$, decreasing $\Gt > 1$ ({\bf spherical-curvature} term dominant)} \\ 
         \multicolumn{2}{c}{} \\ \hline \\
   $\theta$, $\Gamma$, $\calD$, $\ap$ same as for Tail (above)  &  $\theta \simeq \theta_c (t/\tp)^{1/2}$ \\
                                                                     &  $\Gamma = \Gamma_c (t/\tp)^{-g/2}$ \\
                                                                     &  $\calD \sim t^{g/2-1}$  \\ 
   ($t = \tp$): $\ap = \onehalf(2+\beta + i + e\beta)$          &  $\ap = \onehalf[(2-g)(2+\beta)+g(i+e\beta)]$  \\
   $\tpp \simeq 5.4\, (\ln \Gctc)^{2/n} (\Gctc)^2 \to \simeq 5.4\, (\ln \Gctc)^{2/n} \tp$   &  
                                           $\tpp = 2\,(\Gctc)^{2g/(g-1)} \to \simeq  2\,(\Gctc)^{2/(g-1)} \tp$ \\
       \\ \hline \multicolumn{2}{c}{} \\
 \multicolumn{2}{c}{4. POST-PLATEAU from ENVELOPE, $\tpp < t$, decreasing $\Gt < 1$ ({\bf angular-structure} term dominant)} \\ 
         \multicolumn{2}{c}{} \\ \hline \\
   $\theta = \theta_c (\frac{n}{2} \ln t/\to)^{1/n}$             &  $\theta = \theta_c (t/\to)^{1/2g}$ \\
   $\Gamma = \Gamma_c (t/\to)^{-1/2}$                            &  $\Gamma$ same as for $\n$-Exponential (left) \\
   $\calD \sim t^{-1/2}$                                         &  $\calD$ same as for $\n$-Exponential (left) \\
   $\app = \onehalf(i+e\beta+\beta)+2+(1-2/n)/\ln(t/\to)$        &  $\app = \onehalf(i+e\beta+\beta)+2 -1/g$ \\
       \\ \hline \multicolumn{2}{c}{} \\
\end{tabular}
}
\end{table*}

\begin{figure}
\centerline{\includegraphics[width=12cm,height=14cm]{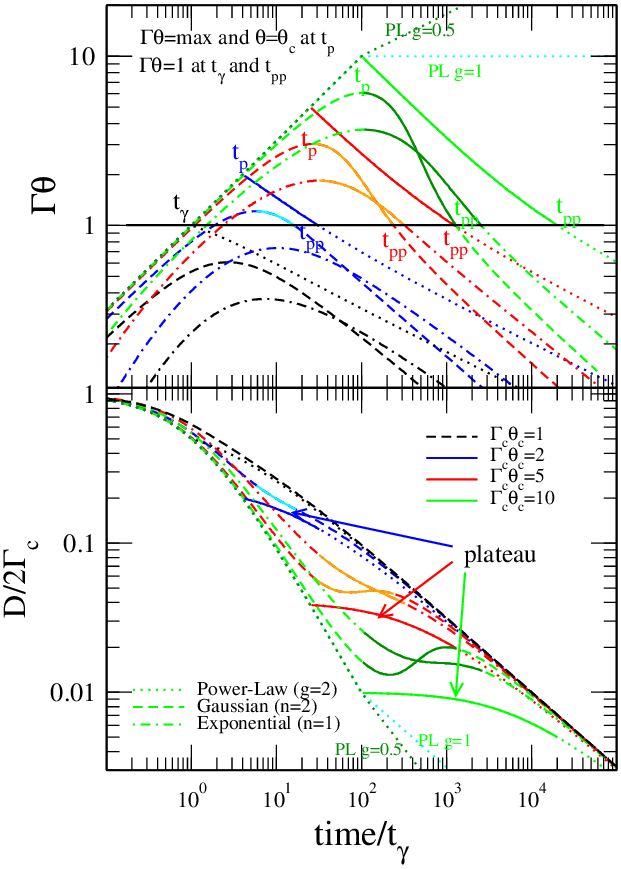}}
\figcaption{ \normalsize
{\sc Upper panel}: Evolution of $\Gamma\theta$ on the outflow surface, determined by solving the equation for photon kinematics
    $\theta (t)$ for each kind of angular structure: 
  "Power-Law" -- $\Gamma = \Gamma_c \thtc^{-2}$ (dotted line),
  "Exponential" -- $\Gamma = \Gamma_c \exp\{-\thtc\}$ (dashed), 
  "Gaussian" -- $\Gamma = \Gamma_c \exp\{-\onehalf \thtc^2\}$ (dot-dashed), 
   and for four outflow parameters $\Gctc$ (color-coded).
  The photon kinematics and the emission relativistic beaming depend only on $\Gctc$, and not on the separate values of those 
   Core parameters (on-axis Lorentz factor $\Gamma_c$, Core angular size $\theta_c$).
  The value of $\Gamma \theta$ relative to unity and its evolution {\bf define analytically the GRB and afterglow phases} 
  (\S\ref{Exp} and \S\ref{PL}):
 i) {\bf GRB} prompt defined by an {\sl increasing $\Gamma \theta < 1$} (emission is beamed toward the observer), 
    lasts until $\tg$ when $\Gamma \theta = 1$ (for the first time), is {\sl produced by the Core} ($\theta < \Gamma_c^{-1} < \theta_c$); 
 ii) GRB {\bf TAIL} defined by an {\sl increasing $\Gamma \theta > 1$} (emission is debeamed), lasts until $\tp \simeq (\Gctc)^2 \tg$ 
    when $\Gamma \theta$ reaches its maximum value ($\siml \Gctc$) at $\theta \simeq \theta_c$, is also {\sl produced by the Core} 
    ($\Gamma_c^{-1} < \theta < \theta_c$);
 iii) afterglow {\bf PLATEAU} (shown with solid lines, off-colors used for $n=1,2$-Exponential outflows) defined by
    a {\sl decreasing $\Gamma \theta > 1$} (observer still outside the $\Gamma^{-1}$ cone of relativistically-enhanced emission), 
    lasts until $\tpp$ when $\Gamma \theta = 1$ (for a second time), is {\sl produced by the Envelope} ($\theta_c < \theta$). 
    Note that, for same Core parameter $\Gctc$, Power-Law outflows yield afterglow Plateaus of longer duration than $\n$-Exponential 
    outflows;
 iv) afterglow {\bf POST-PLATEAU} defined by a {\sl decreasing $\Gamma \theta < 1$} (observer re-enters that cone of beamed emission), 
     is also {\sl produced by the Envelope}.
  For Power-Law outflows, $\Gamma \theta > 1$ only if $\Gctc > 1$; 
  for $\n$-Exponential outflows ($\Gamma = \Gamma_c \exp\{-(1/n)\thtc^n\}$), $\Gamma \theta > 1$ only if $\Gctc > e^{1/n}$. 
  {\sl If these conditions are not satisfied} (black lines), {\sl the GRB Tail and the afterglow Plateau do not exist} according to above 
   definitions (and also because $\tg = \tp \siml \tpp$), and {\sl the prompt GRB is followed by an afterglow post-Plateau/normal-decay}.
  For {\bf Power-Law} outflows of exponent $g \leq 1$ (labeled "PL g=0.5,1"), $\Gamma \theta > 1$ continues to increase from Core to Envelope 
   (i.e. after $\theta = \theta_c$), thus an afterglow phase, defined by a decreasing $\Gamma \theta$, does not exist. 
  However, because this phase follows the prompt one, because its emission originates in the Envelope, and because $\Gt > 1$, 
  it could be called a "pseudo-Plateau".
  For {\bf $\n$-Exponential} outflows, the Lorentz factor in the Envelope decreases sufficiently fast for any exponent $n$ and, 
    if $\Gctc > e^{1/n}$ (i.e. the GRB has a Tail), then $\Gamma \theta > 1$ always decreases below unity after the burst. 
  Therefore, {\sl $\n$-Exponential outflows that yield GRB Tails} (i.e. for $\Gctc > e^{1/n}$) {\sl always produce an afterglow Plateau 
 and a post-Plateau}.
 {\sc Lower panel}: Evolution of the Doppler factor $\calD$, which is dominant in setting the afterglow light-curve,
  and which motivates "visually" the above analytical definitions of GRB/afterglow phases.
  The Plateau lasts longer for a larger Core parameter $\Gctc$ and are more prominent (flatter or even rising) for a larger structural 
 parameter $g$ or $n$. {\bf For the same structural parameters $\Gctc$ and $n/g$, Power-Law outflow yield Plateaus with a larger dynamical range
 than $\n$-Exponential outflows}.
}
\end{figure}

\begin{figure}
\centerline{\includegraphics[width=12cm,height=14cm]{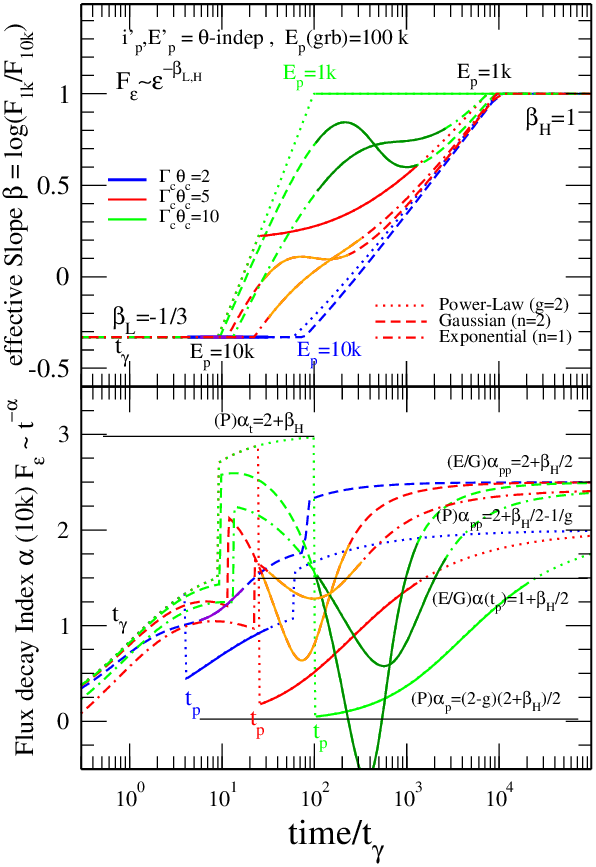}}
\figcaption{ \normalsize
{\sc Upper panel}: Evolution (mostly {\sl softening}) of the effective slope $\beta$ at 1-10 keV for the angular structures and 
 Core parameters of Figure 1, with same style and color-coding, 
 for a broken power-law emission spectrum of (observer-frame) peak-energy $\Ep (t < \tg) = 100$ keV during the GRB,
 and low- and high-energy spectral slopes $\bl = -1/3$ (optically-thin synchrotron/inverse-Compton emission) and $\bh = 1$ (typical value). 
 All three quantities assumed to be {\sl constant/angle-independent} over the emitting surface.
 This effective slope is defined by the ratio of the 1 keV and 10 keV energy flux densities and should be equal to the XRT 
 effective slope (calculated from the hard-to-soft energy hardness-ratio of count-rates) when the spectral peak-energy $\Ep$ 
 is outside the 1-10 keV band, but only an approximate of it in the opposite case (the spectral evolution should be 
 quantitatively the same for both slopes). 
 Evidently, $\beta = \bl$ when $\Ep > 10$ keV and $\beta = \bh$ when $\Ep < 1$ keV.
 Epochs when $\Ep$ crosses the 1-10 keV boundaries are indicated.
{\sc Lower panel}: 
 Evolution of the power-law decay index $\alpha$ of the afterglow flux at 10 keV, {\sl consistent with the
 analytical asymptotic expectations} for the GRB Tail (subscript "t"), Plateau ("p"), and post-Plateau ("pp"), and for a 
 (P)ower-Law (dotted lines), an $n=1$ (E)xponential (dot-dashed), and a (G)aussian (dashed, $n=2$ Exponential) outflow. 
 The comoving-frame spectral peak-energy $E'_p$ and peak-intensity are assumed constant (uniform) over the outflow surface, 
 thus {\sl these decay indices are due solely to the evolution of the Doppler factor $\calD$} (Figure 1). 
 The $\Ep$ decreasing below 10 keV (upper panel) yields a steepening of the flux decay because the emission spectrum is harder
 above $\Ep$ than below it ($\bh < \bl$ in the $F_\eps \sim \eps^{-\beta}$ definition).
 The analytical asymptotic indices $\alpha$ depend only on the high-energy slope $\bh$ because $\Ep < 10$ keV at those times. 
 Following lines of same coding shows that, well after $\tpp$, {\sl the post-Plateau decay indices for an Exponential and Gaussian outflow} 
 approach asymptotically the same index $\app$ (which is true for any exponent $n$, see Table 1), and {\sl are larger than the index 
 for a Power-Law outflow} by $1/g\simeq 1/2$.
 Following lines of same color shows that the {\bf Plateaus of Power-Law outflows} (with $g=2$) {\bf last longer than for $\n$-Exponential 
 outflows} (with $n \in [1,2]$).
}
\end{figure}

\begin{figure}
\centerline{\includegraphics[width=13cm,height=11cm]{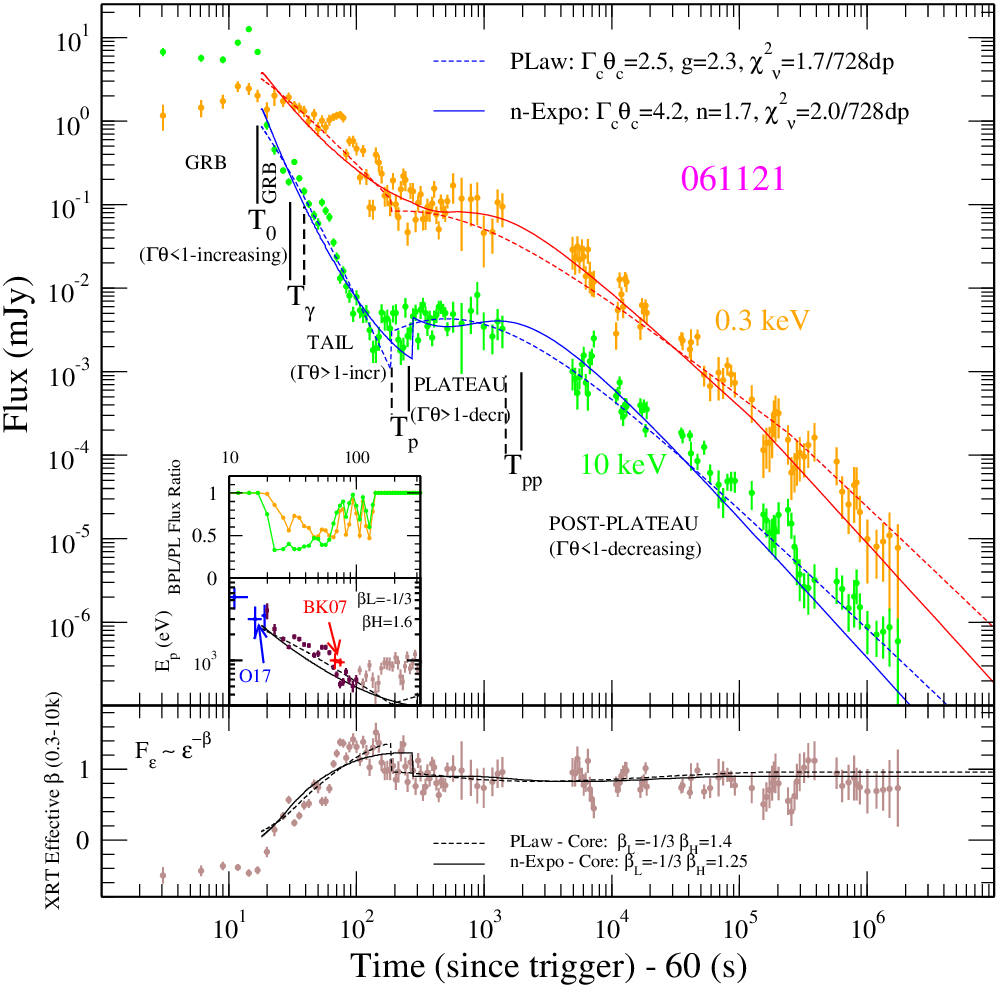}}
\figcaption{ \normalsize
 {\sc Upper panel}: $LAE$ model best-fits for outflows with a $\n$-Exponential (solid lines) and a $g$-Power-Law (dashed) angular structure 
  found by $\chi^2$-minimization to the 0.3 keV (green symbols) and 10 keV (orange) XRT fluxes of GRB 061121 -- ({\sl a GRB with Tail 
  and an afterglow with Plateau} ({\bf type I}) extending over {\bf 1 dex}). 
  Model fitting is done using all XRT measurements; for clarity, this figure shows fewer data, obtained by rebinning measurements within 
  $\delta t = 0.03 t$ (as for the rest of Figures) and by calculating error-weighted averages and uncertainties.  
   The 0.3 keV and 10 keV light-curves are calculated from the bolometric 0.3-10 keV fluxes and effective slopes found at the XRT repository,
  as described in \S\ref{XRT},
  using a broken power-law spectrum (with the low-energy and high-energy slopes $\bl = -1/3$ and $\bh = 1.6$ being the effective slopes 
  (lower panel) measured at the beginning (when $\Ep > 10$ keV) and end (when $\Ep < 0.3$ keV) of the GRB Tail)
	until the Plateau beginning at $\Tp \sim 200$, and a single power-law spectrum (with the slope measured by XRT) after that.
   Basic parameters ($\Gctc$, $g$, $n$) and temporal milestones/light-curve breaks ($\To$,$\Tg$,$\Tp$,$\Tpp$), as well as the four emission 
   phases (and their definitions) are indicated. 
  Other best-fit parameters are:
  low-energy (below $\Ep$) spectral slope $\bl = -1/3$ (fixed at the value expected for optically-thin synchrotron/inverse-Compton emission), 
  high-energy (above $\Ep$) slope $\bh \simeq 1.3$ for the Core ($\theta < \theta_c$) emission 
  (i.e. for the prompt GRB and Tail) and $\bh \simeq 1.0$ for the Envelope ($\theta > \theta_c$) emission (i.e. the afterglow). 
   A {\sl slope $\bh$ that is discontinuous} at the Core-Envelope boundary {\sl is required by the hardening of the effective slope}
  $\beta$ measured by XRT around $\Tp$, at the end of the GRB Tail and the beginning of the afterglow Plateau (lower panel).
  All other model parameters are continuous at that boundary.
  Based on their respective $\chi^2$, {\sl the Power-Law outflow provides a better fit than the Exponential surface} although that is not
  immediately evident to the "$\chi$-by-eye" observer.
   {\sc Upper inset}: ratio of fluxes calculated from the XRT data using a broken power-law spectrum to those for a pure power-law spectrum 
   of XRT's effective slope $\beta$. 
  That ratio is below unity only when the peak-energy $\Ep$ of the broken power-law spectrum is in the XRT 0.3-10 keV window, 
  i.e. when there is a spectral evolution of the XRT effective slope in the lower panel, which happens only during the GRB Tail. 
  {\sc Lower inset} shows the peak-energy $\Ep$ of the broken power-law (synchrotron) spectrum (same line-coding as model light-curves). 
  (Light symbols show the peak-energy $\Ep$ (in the XRT band) that accommodates the effective slope $\beta$ measured by XRT {\sl after}
  the GRB Tail, with $\Ep$ calculated for the slopes $\bl$ and $\bh$ measured during the GRB Tail. However, the non-evolving slope $\beta$ 
  measured during the Plateau indicates that the real peak-energy $\Ep$ is {\sl not} in the XRT window, i.e. extending the calculation 
  of $\Ep$ to the Plateau was a useless excercise, done just for the fun of it.) 
  Red (BK07) and blue (O17) symbols indicate the peak energies of $\eps F_\eps$ measured by Butler \& Kocevski (2007) and by Oganesyan 
  et al (2017), respectively, by fitting the XRT and BAT spectra with a broken power-law. 
   The $\Ep$ values derived here by matching the XRT effective slope $\beta$ {\sl are consistent} with those measurements.
 {\sc Lower panel}: XRT effective slope $\beta$ at 0.3-10 keV from the XRT repository. Lines show the effective slope $\beta$ of the best-fit model.
}
\end{figure}

\begin{figure}
\centerline{\includegraphics[width=13cm,height=11cm]{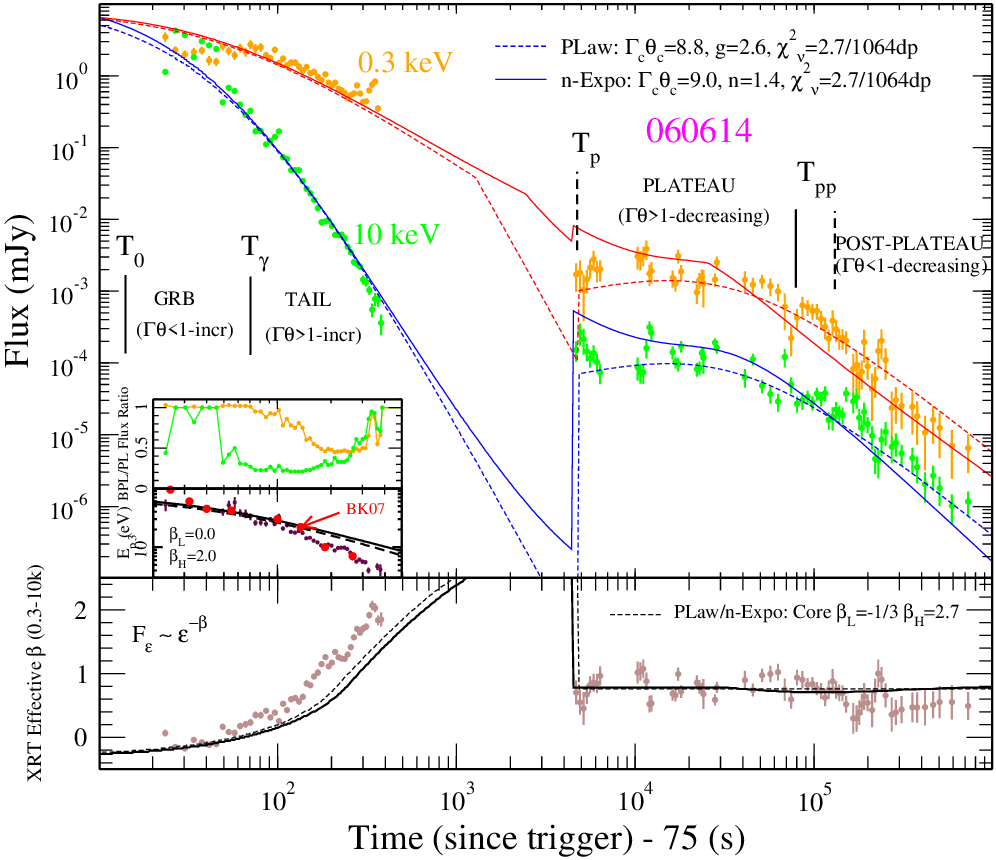}}
\figcaption{ \normalsize
 {(\bf upper panel)} $LAE$ model best-fits for outflows with a $\n$-Exponential (solid lines) and a $g$-Power-Law (dashed) angular structure 
 to the XRT light-curves of GRB 060614 ({\sl a GRB with Tail and an afterglow with Plateau} ({\bf type I}) extending over {\bf 1.5 dex}).
  Lacking a measurement of the Plateau beginning epoch $\Tp$ means that the remaining temporal milestones $\To$ (pulse beginning), 
  $\Tg$ (GRB end), and Plateau end ($\Tpp$) cannot constrain both fundamental model parameters: Core's $\Gctc$ and the angular structure 
  indices $g$ or $n$. Consequently, {\sl Power-Law and n-Exponential outflows accommodate equally well} the X-ray emission of this burst. 
  Due to the lack of a strong constraint on $\Gctc$, the best-fitting routine pushes that parameter to its maximal value 
  (thus $\Tp$ is set to just before the first Plateau measurement at 5 ks), so that the Plateau-end epoch $\Tpp$ is maximized.
  The lower inset shows that the peak-energy $\Ep$ calculated from the XRT effective slope $\beta$ (lower panel) and the assumed
  low-and high-energy slopes $\bl=0.0$ and $\bh=2.0$ of the Core's emission spectrum (which are the effective slopes measured at 
  the beginning and end of the prompt phase) are consistent with the spectral-fit measurements of Butler \& Kocevski (2007) (BK07). 
}
\end{figure}

\begin{figure}
\centerline{\includegraphics[width=13cm,height=11cm]{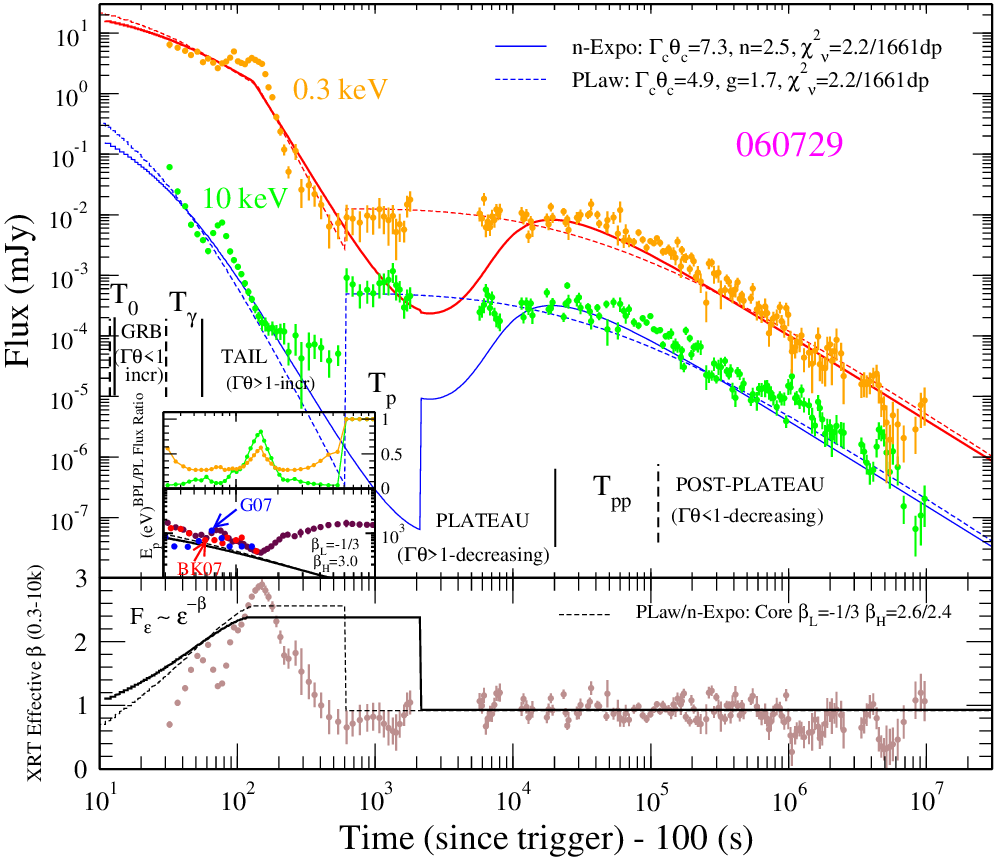}}
\figcaption{ \normalsize
  $LAE$ model best-fits for outflows with a $\n$-Exponential (solid lines) and a $g$-Power-Law (dashed) angular structure to the 
  XRT light-curves of GRB 060729 -- ({\sl a GRB with Tail and an afterglow with Plateau} ({\bf type I}) extending over {\bf 2 dex}).
   The 0.3 keV and 10 keV measured fluxes are calculated using a broken power-law spectrum until 2 ks, i.e. beyond the epoch (250 s)
  when the softest XRT slope is reached and beyond the end of the GRB tail (600 s) of the 0.3-10 keV XRT flux. 
   Restricting the use of the broken power-law spectrum (for flux calculation from the XRT data) to those earlier epochs changes 
  little the resulting best-fit, which is equally good to the 0.3 keV Tail and just slightly worse to the 10 keV Tail because 
  $\chi^2$ is not sensitive to an extra under-prediction.
   The prompt emission displays a {\sl challenging feature}: slightly decoupled Tails, with the 0.3 keV Tail light-curve 
  delayed relative to the 10 keV's. 
   The lower panel shows that the reason for that delay and decoupling is the strong spectral hardening of the Tail emission from
  the outflow Core after 250 s, which is more likely due to a {\sl hardening of the high-energy slope} $\bh$ for most 
  of the GRB Tail than to an increase of the peak-energy $\Ep$ (\S\ref{NumFits}). 
   An evolving spectral slope is not a feature currently included in the $LAE$ model, thus this model cannot account for any strongly
  decoupled behavior of GRB Tails.
   However, that model deficiency cannot shorten the Plateau duration and the conclusion that $\n$-Exponential outflows fail 
  to account for this Plateau (which is longer than 1.5 dex), stands (\S\ref{numerics}).
  Red symbols (BK07) show the peak-energy $\Ep$ measured by Butler \& Kocevski (2007) by fitting the XRT+BAT spectrum with a broken 
  power-law, blue symbols (G07) show the cut-off energy measured by Grupe et al (2007) using a power-law with an exponential cutoff.
}
\end{figure}

\begin{figure}
\centerline{\includegraphics[width=13cm,height=11cm]{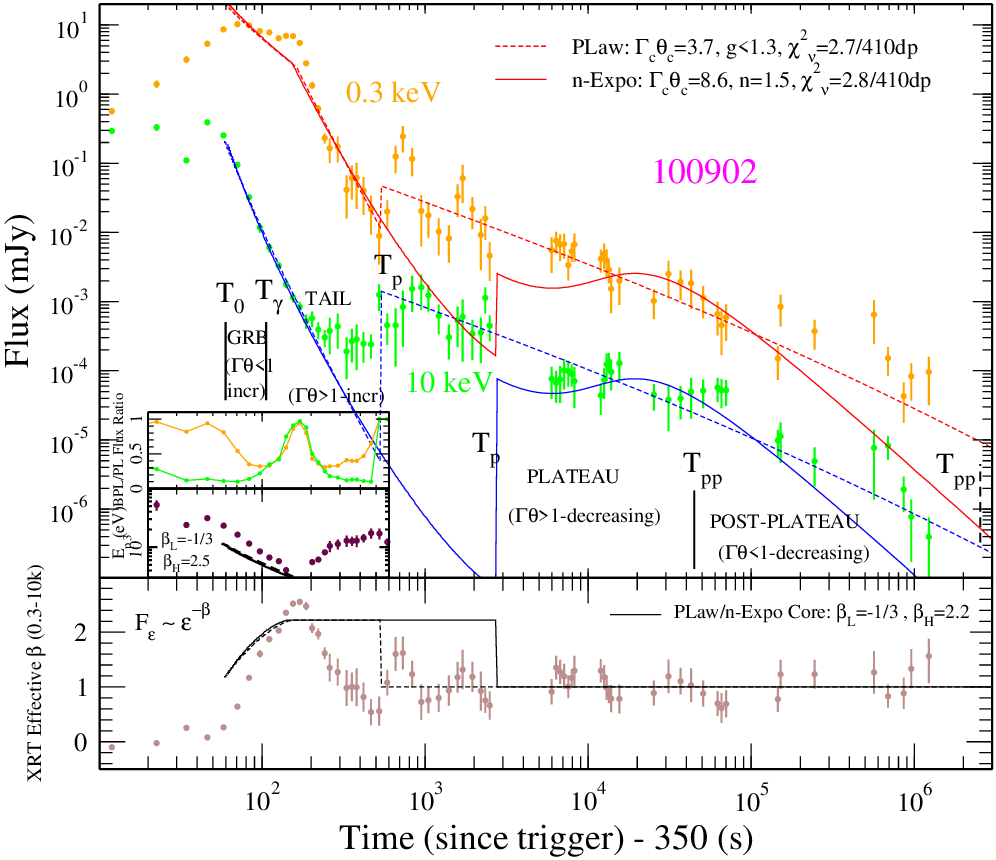}}
\figcaption{ \normalsize
  $LAE$ model best-fits for outflows with a $\n$-Exponential (solid lines) and a $g$-Power-Law (dashed) angular structure to the XRT 
 light-curves of GRB 100902 -- ({\sl a GRB with Tail and an afterglow with Plateau} ({\bf type I}) extending over {\bf 3 dex} and of
 (borderline) power-law flux-decay index $\ap = 0.75$. 
 GRB afterglows of type I are naturally explained by $\n$-Exponential outflows, for which the Lorentz factor $\Gamma$ decreases in the Envelope 
 sufficiently fast to lead to a decreasing $\Gt > 1$ (i.e. to a Plateau), followed by a decreasing $\Gt < 1$ (i.e. to a post-Plateau).
  For Power-Law outflows, type I afterglows with such long Plateaus require either a large parameter $\Gctc$ or, better yet, a structural 
 index $g \simg 1$ (Equation \ref{tppPLaw}) ($g=1.3$ shown here).
  Alternatively, GRB 100902 could be a {\sl GRB with Tail and an Afterglow with a pseudo-Plateau} ({\bf type II}), which requires $g < 1$ 
 for a Power-Law structure, but which could pose a problem for $\n$-Exponential outflows, as they should always yield two afterglow phases. 
  Similar to GRB 060729, the 0.3 keV and 10 keV light-curves are decoupled toward the Tail end, which is the result of a strong spectral hardening 
  occuring at that time (lower inset), and which indicates a hardening of the high-energy slope $\bh$ of the X-ray continuum.
  This burst illustrates {\sl the intrinsic difficulty of the $\n$-Exponential structure in accommodating Plateaus longer than 2 dex}.
}
\end{figure}

\begin{figure}
\centerline{\includegraphics[width=13cm,height=11cm]{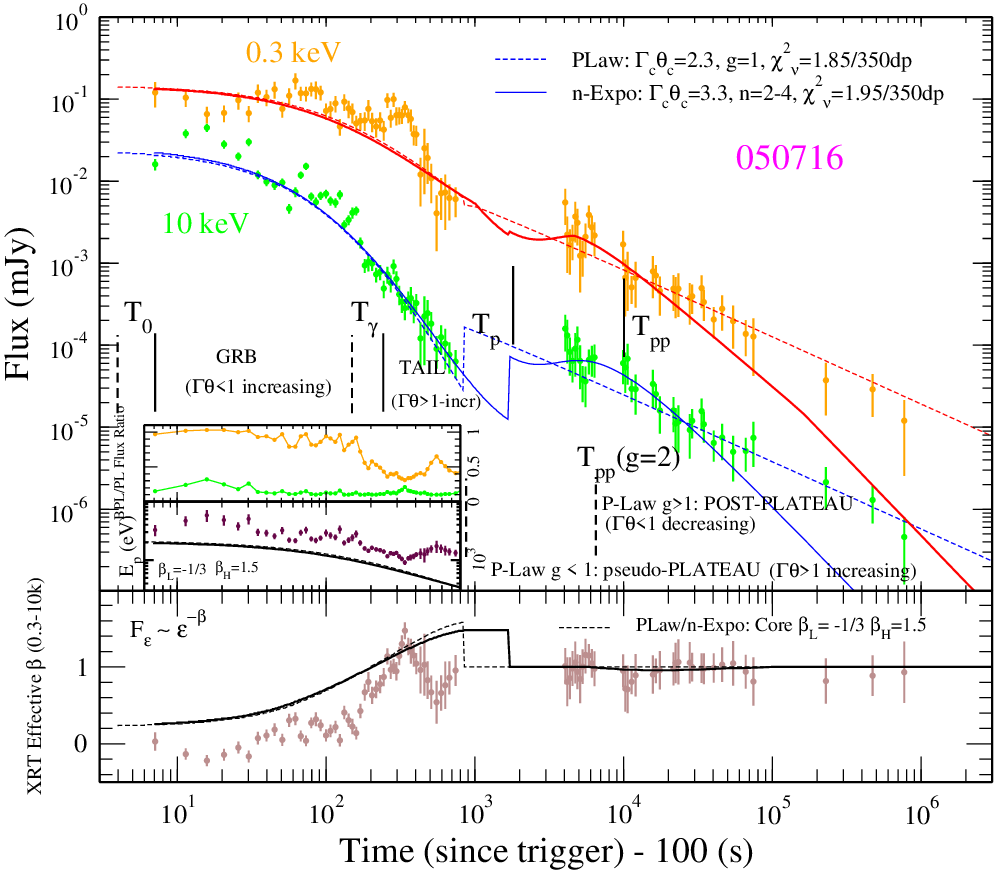}}
\figcaption{ \normalsize
  $LAE$ model best-fits for outflows with a $\n$-Exponential (solid lines) and a $g$-Power-Law (dashed) angular structure to the XRT
 light-curves of GRB 050716 -- ({\sl a GRB with Tail and an afterglow with a pseudo-Plateau} ({\bf type II}) of power-law flux-decay index 
 $\ap = 0.9$ (a little too steept for a Plateau).
  Due to the Earth's occultation, a Plateau shorter than 4 ks may have been missed, in which case GRB 050716 would be a ({\sl a GRB with 
  Tail and an afterglow with Plateau} ({\bf type I}) extending over less than {\bf 1 dex}, followed by a post-Plateau of index $\app=1.0$.
  This ambiguity about the Plateau existence implies that the structural parameter $g$ of the best-fit Power-Law outflow is uncertain: 
 equally good best-fits can be obtained with $g < 1$ (for which the Envelope emission satisfies $\Gt > 1$ but is increasing -- Figure 1, 
 thus the afterglow emission satisfies only half of the Plateau definition, and can be called a "pseudo-Plateau"), and with $g \in (1,3)$ 
  which leads to a Plateau (satisfying its full definition: a decreasing $\Gt > 1$) and a post-Plateau (defined by a decreasing $\Gt < 1$).
  In contrast, the Lorentz factor decreases sufficiently fast in the Envelope of an $\n$-Exponential outflow and leads to a decreasing $\Gt$,
 i.e. a Plateau and post-Plateau are guaranteed to exist for this angular structure. 
  While discriminating among these two types of structures based on that, for $\Gctc > 1$, an $\n$-Exponential outflow cannot produce a single 
 afterglow phase (like a Power-Law outflow does for $g < 1$) will always be hampered by that short Plateaus can "hide" behind the Earth,
 {\sl the faster post-Plateau flux-decay characteristic of $\n$-Exponential outflows} (which cannot be mitigated by adequate evolutions
 of the spectral characteristics $i'_p$ and $E'_p$ because those evolutions are constrained by the Tail flux) {\sl leads to a poorer best-fit 
 than for a Power-Law outflow}.
}
\end{figure}

\begin{figure}
\centerline{\includegraphics[width=13cm,height=11cm]{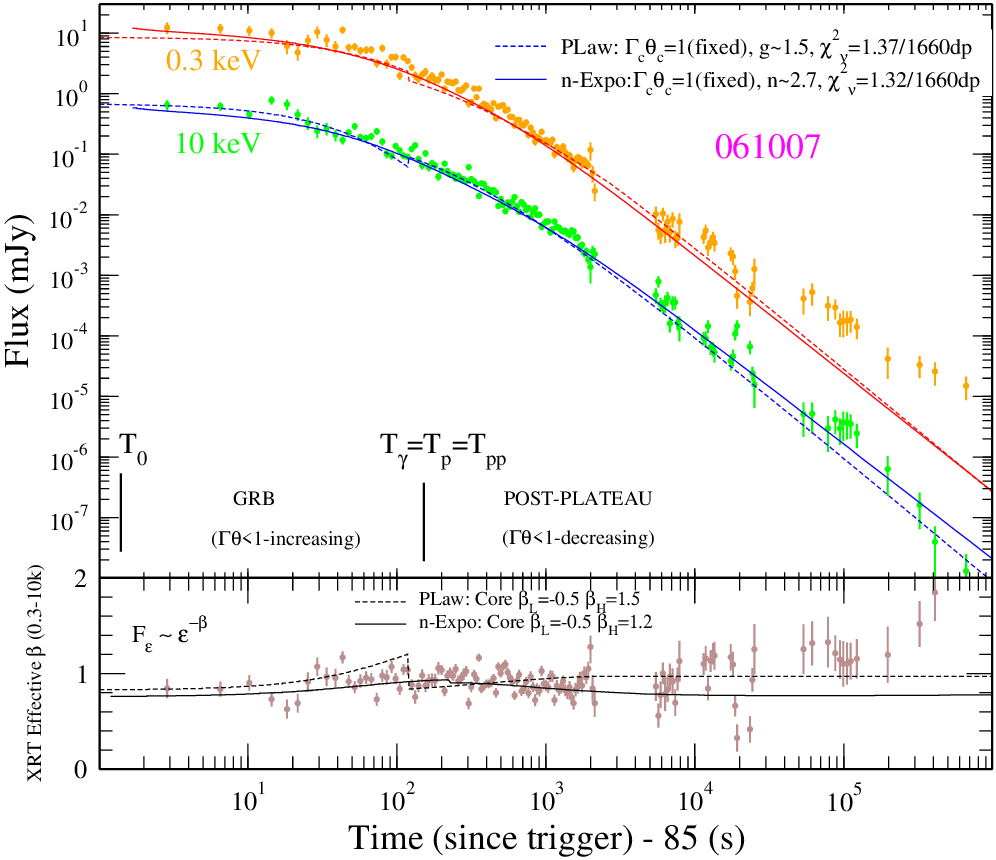}}
\figcaption{ \normalsize
  $LAE$ model best-fits for outflows with a $\n$-Exponential (solid lines) and a $g$-Power-Law (dashed) angular structure to the XRT 
 light-curves of GRB 061007 -- ({\sl a GRB without Tail and an afterglow without a Plateau} ({\bf type III}) (BAT data prior to 85 s also 
 do not display a Tail). The lack of a GRB Tail and of an afterglow Plateau, indicate that $\Gctc \leq 1$ for a Power-Law angular structure 
 and $\Gctc \leq e^{1/n}$ for a $\n$-Exponential outflow, for which $\Gt < 1$ at all times, as shown in Figure 1. 
 Here, the Core parameter $\Gctc$ was fixed at unity, but comprably good fits can be obtained for other sub-unity values. 
 The angular structure exponents $g$ and $n$ are uncertain by about 0.5.
 The two types of angular structures accommodate equally well the light-curves because the lack of a Plateau reduces the number of 
 observational constraints on the angular structure. 
  Thus, {\sl GRBs without Tails followed by afterglows without Plateaus cannot constrain the outflow structure.}
 The spectral softening after 30 ks could be due to the evolution of the high-energy spectral slope $\bh$ of the Envelope emission,
 a feature that is not included in the $LAE$ model, where $\bh$ is assumed constant.
}
\end{figure}


\end{document}